\documentclass{ws-ijmpb}

\usepackage{cite,latexsym,amssymb,graphicx}

\begin{document}

\markboth{P. Cain, R. A. R\"omer} {Real space RG approach to
  the integer QH effect}
%
\catchline{}{}{}{}{}
%

\title{Real-space renormalization-group approach\\ to the integer quantum
  Hall effect}

\author{Philipp Cain}
\address{Institut f\"ur Physik,
Technische Universit\"at, 09107 Chemnitz, Germany\\
cain@physik.tu-chemnitz.de}

\author{Rudolf A.\ R\"omer} \address{Department of Physics and Centre
  for Scientific Computing, University of Warwick, Coventry CV4 7AL,
  United Kingdom\\
 r.roemer@warwick.ac.uk}

\maketitle

\begin{history}
\received{January 2005}
\revised{$Revision: 1.4 $}
\end{history}
\begin{abstract}
  We review recent results based on an application of the real-space
  renormalization group (RG) approach to a network model for the integer
  quantum Hall (QH) transition. We demonstrate that this RG approach
  reproduces the critical distribution of the power transmission
  coefficients, i.e., two-terminal conductances, $P_{\rm c}(G)$, with
  very high accuracy. The RG flow of $P(G)$ at energies away from the
  transition yields a value of the critical exponent $\nu$ that agrees
  with most accurate large-size lattice simulations. A description of
  how to obtain other relevant transport coefficients such as $R_{\rm
    L}$ and $R_{\rm H}$ is given.
  From the non-trivial fixed point of the RG flow we extract the
  critical level-spacing distribution (LSD). This distribution is close,
  but distinctively different from the earlier large-scale simulations.
  We find that the LSD obeys scaling behavior around the QH transition
  with $\nu= 2.37\pm 0.02$. Away from the transition it crosses over
  towards the Poisson distribution.
  We next investigate the plateau-to-insulator transition at strong
  magnetic field. For a fully quantum coherent situation, we find a
  quantized Hall insulator with $R_{\rm H}\approx h/e^2$ up to $R_{\rm
    L}\sim 20 h/e^2$ when interpreting the results in terms of most
  probable value of the distribution function $P(R_{\rm H})$. Upon
  further increasing $R_{\rm L}\rightarrow\infty$, the Hall insulator
  with diverging Hall resistance $R_{\rm H}\propto R_{\rm L}^{\kappa}$
  is seen. The crossover between these two regimes depends on the
  precise nature of the averaging procedure for the distributions
  $P(R_{\rm L})$ and $P(R_{\rm H})$.
  We also study the effect of long-ranged inhomogeneities on the
  critical properties of the QH transition.  Inhomogeneities are modeled
  by a smooth random potential with a correlator which falls off with
  distance as a power law $r^{-\alpha}$. Similar to the classical
  percolation, we observe an enhancement of $\nu$ with decreasing
  $\alpha$.
  These results exemplify the surprising fact that a small RG unit,
  containing only five nodes, accurately captures most of the
  correlations responsible for the localization-delocalization
  transition.
\end{abstract}
\keywords{integer quantum Hall effect; real-space renormalization group
  approach; network models; energy-level statistics; macroscopic
  inhomogeneities; quantum Hall insulator.}

\section{Introduction}
\label{chap-intro}

Measuring the resistance of a two-dimensional (2D) electron gas at
very low temperature and subject to a strong perpendicular
magnetic field $B$ reveals a striking macroscopic quantum
phenomenon \cite{KliDP80}:  the Hall resistance $R_{\rm H}$ shows
very precise plateaus at $(1/N) h/e^2$ where $N$ is an integer
number. In the same region the vanishing longitudinal resistance
$R_{\rm L}$ indicates a dissipationless flow of current.
Quantization can also be observed at fractional values of $N$
\cite{TsuSG82}, which leads to the distinction between the fractional
and the integer quantum Hall (QH) effect. Despite the similarities
regarding the experimental observations the theoretical description of
both effects differs considerably. The integer QH effect can be
explained reasonably well within a non-interacting electron picture,
while interactions play a fundamental role in the fractional QH effect
\cite{ChaP95,Yos02}.

\begin{figure}[tbh]
\centerline{\includegraphics[width=0.6\columnwidth]{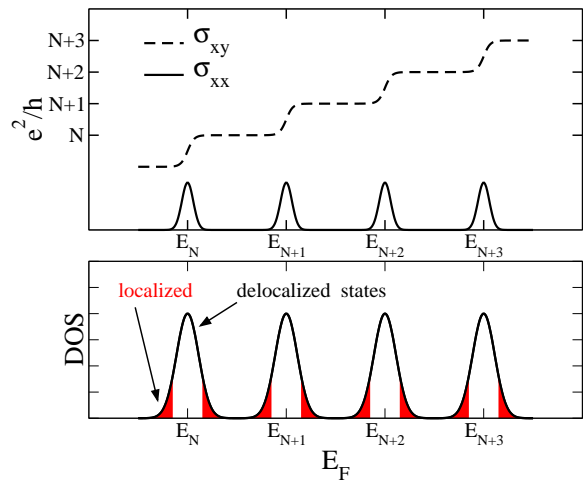}}
\caption{\label{fig-qhe-dos}
  Schematic behavior of longitudinal ($\sigma_{xx}$) and Hall
  ($\sigma_{xy}$) conductivities as well as the density-of-states (DOS)
  for the integer QH effect.  $E_{\rm F}$ denotes the Fermi energy and
  the energy of the Landau levels is $E= \hbar \omega_{\rm c} (N -
  1/2)$, $N= 1, 2, \ldots$; $\omega_{\rm c}= e B/m$ the cyclotron
  frequency.  The width of the region of extended states at the band
  centers of the disorder-broadened Landau levels is finite due to
  finite temperatures and finite system sizes. }
\end{figure}

The integer QH effect can be rationalized at a phenomenological
level as a series of localization-delocalization transitions
\cite{AokA81}. For non-interacting electrons, the density of
states consists of distinct Landau bands. These are
broadened by the intrinsic disorder of the QH samples. Each such
band contains localized states in the tails and extended states in
the center of the band. As the Fermi energy passes through the
extended states, the Hall resistance $R_{\rm H}$ rises to the next
plateau and the longitudinal resistance $R_{\rm L}$ is
non-vanishing. In the range between the bands and for the
localized states the Hall resistance remains fixed at the plateau
value as shown in Fig.\ \ref{fig-qhe-dos}.

Several theoretical approaches have been proposed to explain the
sequence of integer QH transitions in more detail
\cite{AokA81,ChaP95,JanVFH94,Lau81,Pra81,Pru84,ThoKNN82,Yos02}.  For
example, it has been established that the QH transition is a
second-order phase transition \cite{Huc95} where a power-law divergence
of characteristic length scales is observed. The divergence, e.g.\ of
the localization length $\lambda$ of the wave functions, $\lambda\propto
\epsilon^{-\nu}$, where $\epsilon$ corresponds to the distance from the
transition, can be quantified by the exponent $\nu$. Its value is a
universal quantity, independent from microscopic details of electron
motion, disorder realization and even the chosen material.

The Chalker-Coddington (CC) model is a convenient representation of the
QH situation in the fully quantum coherent regime at low
temperature and strong magnetic field \cite{ChaC88}. The model
describes a single Landau band and thus contains only one QH
transition. Electrons are treated semiclassically as
non-interacting particles moving in a smoothly varying 2D disorder
potential under a strong perpendicular magnetic field. The
percolation of the electron through the sample is then given by
motion along equipotentials interrupted by scattering events at
saddle points (SPs) of the potential. For theoretical
investigations, equipotentials and SPs are mapped on links and
nodes of a 2D network, respectively, and large networks have to be
studied in order to reduce finite-size effects
\cite{BatS96,ChaC88,Huc92,KleM97,LeeWK93,Met98b}.

In the present review, we concentrate on an alternative
approach to the CC model. Namely, we shall introduce a real-space
renormalization-group (RG) approach \cite{AroJS97,GalR97}, based
on mapping a characteristic part of the CC network --- the RG unit
--- consisting of several SPs onto a single super-SP
residing in a new CC-super network. This approach is motivated by
the success of RG techniques in statistical physics and
particularly the study of phase transitions \cite{BinDFN92,Wil83}.
Common to both real-space\cite{Car96} and field-theoretical
\cite{Sal99} RGs is the elimination of irrelevant (either
short-range or short-period) degrees of freedom. Afterwards the
original phase volume is restored by a scale transformation. The
system has now the same structure as the original but the
parameters, e.g.\ coupling constants, are {\it renormalized}.
During the iteration of the above RG steps the parameters approach
their fixed point (FP) values and an analysis of this convergence
allows one to derive the critical properties of the transition.
Large effective system sizes can be achieved in this way for the
CC model. This advantage is counterbalanced by the approximate
description of the network due to the limited size of the RG unit.

The RG procedure provides a natural route to studying the complete
distribution functions of the 2D resistivities defined via
\begin{equation}
{\bf E}=\left(
\begin{array}{cc}
  \rho_{xx} & \rho_{xy}\\
  -\rho_{xy} & \rho_{xx}\\
\end{array}
\right) {\bf j} ,
\end{equation}
where $\rho_{xx}$ is called longitudinal and $\rho_{xy}$ Hall
resistivity. The conductivity tensor is defined as the inverse of
the resistivity tensor which yields
\begin{equation}
\label{eq-qhe-sigma-rho}
\sigma_{xx}=\sigma_{yy}=\frac{\rho_{xx}}{\rho_{xx}^2+\rho_{xy}^2}
\quad \mbox{and} \quad
\sigma_{xy}=-\sigma_{yx}=-\frac{\rho_{xy}}{\rho_{xx}^2+\rho_{xy}^2}.
\end{equation}
We emphasize that in 2D resistances (conductances) and resistivities
(conductivities) are related by a simple, dimensionless geometrical
factor.

As a first application of the real-space RG approach to the CC
model, we study the distribution functions $P$ of two-point
conductance $G$ and the resistances $R_{\rm L}$ and $R_{\rm H}$.
From these, we compute the value of the critical exponent $\nu$.
Furthermore, closing the incoming and outgoing links of the RG
unit and thus quantizing the allowed energies, we can make contact
with previous studies of energy level statistics. A suitable
scaling ansatz again allows the independent estimation of $\nu$.
These results are in good agreement with previous studies and
validate the present RG approach.

A further application is concerned with the possible existence of
the quantized Hall insulator. While it is established that
plateau-plateau and insulator-plateau transitions exhibit the same
critical behavior
\cite{GolWSS93,HilSST98,LanPVP02,ShaTSC97,ShaTSS96,HugNFL94,MurHLC00,PanSTW97,ShaTSB95,ShaTSS97,SchVOW00}
the value of the Hall resistance $R_{\rm H}$ in this insulating
phase is still rather controversial \cite{Shi04}. Various
experiments have found that $R_{\rm H}$ remains very close to its
quantized value $h/e^2$ even deep in the insulating regime when
already $R_{\rm L} \gg h/e^2$
\cite{HilSST98,LanPVP02,PelSCS03,ShaTSC97,ShaTSS96}. This scenario
has been dubbed the {\em quantized Hall insulator}. On the other
hand, theoretical predictions based on quantum coherent models
show that a diverging $R_{\rm H}$ should be expected
\cite{PryA99,ZulS01}. Extending the RG approach to the calculation
of suitable means for $R_{\rm H}$ and $R_{\rm L}$ we show that the
RG approach can in fact reconcile these findings by establishing
that the most-probable value of $R_{\rm H}$ remains rather close
to $h/e^2$ even for large $R_{\rm L}$, but then diverges with the
predicted power-law for $R_{\rm L} > 20 h/e^2$.

Last, the RG approach is also ideally placed to study the changes of the
critical properties due to truly long-ranged,\footnote{Note that
  sometimes the term ``long-ranged disorder'' is also used for a
  disorder that has a finite correlation radius which is larger than the
  magnetic length \cite{EveMPW99}. This is different from the present
  situation.}  so-called macroscopic inhomogeneities. Close to the
transition the localization length $\xi$ becomes sufficiently large.
Then the long-ranged disorder can affect the character of the divergence
of $\xi$. We note that in previous considerations inhomogeneities were
incorporated into the theory through a spatial variation of the {\em
  local} resistivity \cite{CooHHR97,RuzCH96,Shi99,ShiAK98,SimH94}, i.e.,
``inhomogeneously {\em incoherent}''.  Our present approach is able to
retain the full quantum coherence.

The results of this study are presented as follows.
The main tool of this work, the RG approach to the CC model, is content
of section \ref{sec-rgapproach}. Results for resistance and conductance
distributions, the level spacing statistics and the associated estimates
of the critical exponent are reviewed in section \ref{sec-pglsd}. The
extension of the RG approach for the QH insulator and for macroscopic
inhomogeneities are presented in section \ref{sec-rhmacro}. We conclude
in section \ref{sec-concl}.

\section{The quantum RG approach to the CC model}
\label{sec-rgapproach}

\subsection{The Chalker-Coddington network model}
\label{sec-rgapproach-ccmodel}

\begin{figure}[tbh]
\center{\includegraphics[width=0.45\textwidth]{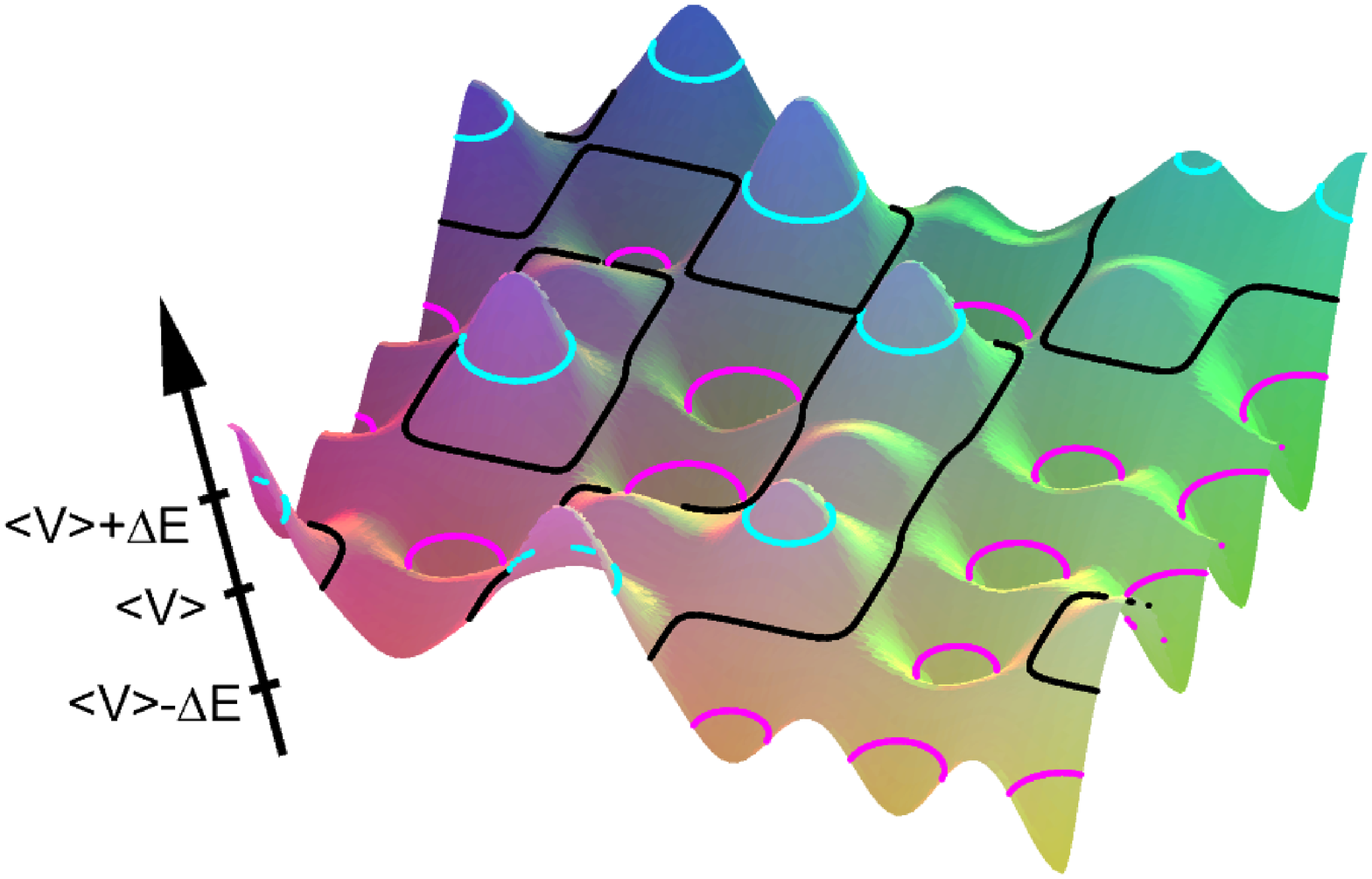}
\quad\includegraphics[width=0.45\textwidth]{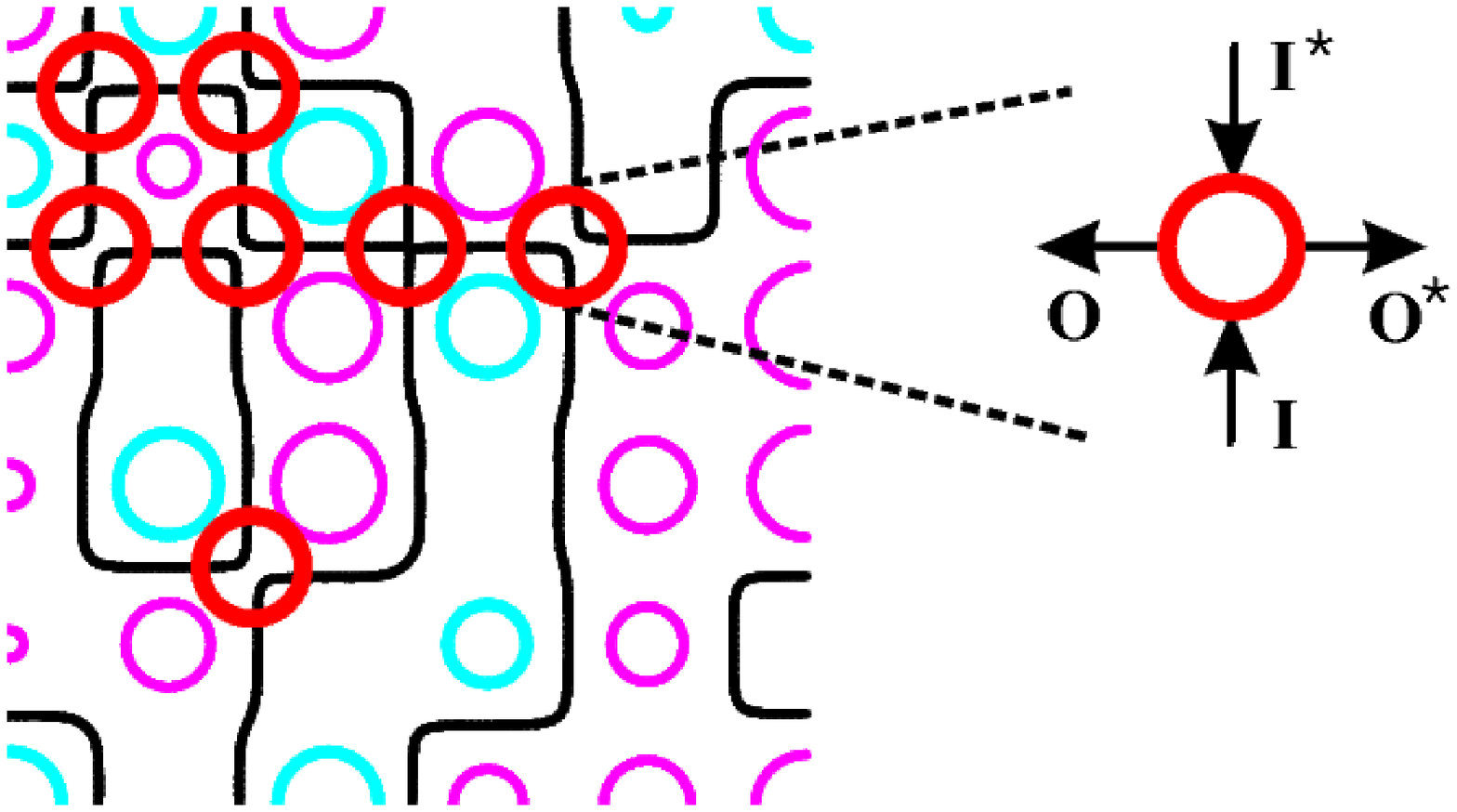}}
\caption{\label{fig-qhe-landscape}\label{fig-qhe-SPnetwork}
  Left: Illustration of a weakly varying random potential $V$ with
  equipotential lines at $\langle V \rangle$ and $\langle V \rangle\pm
  \Delta E$ which allows to separate the electron motion (black orbit)
  in a strong magnetic field $B$ into cyclotron motion and motion
  of the guiding center along equipotentials.
  Right: From these equipotentials at $E=\langle V\rangle$, $\langle V
  \rangle\pm\Delta E$ SPs of the potential can be identified. A single
  SP acts as a scatterer connecting two incoming with two outgoing
  channels.}
\end{figure}

The CC network model \cite{ChaC88} uses a semi-classical approach to
model the integer QH transition and is one of the main ``tools'' for the
quantitative study of the QH transition
\cite{JanMZ99,KagHA95,KagHA97,KleM95,KleM97,KleZ01,LeeC94,LeeCK94,LeeWK93,Met98b,RuzF95,WanLW94}.
It is based on an extension of the high-field model \cite{Ior82}.  In
order to include the previously mentioned localization-delocalization
scenario the classical high-field model relies on two basic
prerequisites.  First, the 2D sample is penetrated by a very strong
perpendicular magnetic field and second the electrons are
non-interacting and move in a smoothly varying 2D potential energy
landscape $V({\bf r})$ illustrated in Fig.\ \ref{fig-qhe-landscape}.
As a result the magnetic field $B$ forces an electron onto a cyclotron
motion with radius $l_{\rm B}$ much smaller than the potential
fluctuations.  Thus the electron motion can be separated into cyclotron
motion and a motion of the guiding center along equipotentials of the
energy landscape \cite{Ior82}.  The cyclotron motion leads to the
required quantization into discrete Landau levels whereas its influence
on the electron motion in the potential can be neglected. Besides the
fundamental assumption about the difference of length scales the model
therefore contains no further dependence on $B$.  The quantum effects of
the electron transport through the sample are only determined by the
height of the SPs in the potential energy landscape.  This picture is
related to a classical bond percolation problem on a square lattice
\cite{StaA95} when mapping SPs onto bonds.  A bond is connecting only
when the potential of the corresponding SP equals the potential energy
$\varepsilon$ of the electron. From percolation theory \cite{StaA95}
follows that an infinite system is conducting only at a single
$\varepsilon=\langle V \rangle$.  Consequently, the high-field model
describes only a single QH transition. It provides a qualitative
understanding of the existence of an localization-delocalization
transition and thus the quantized plateaus in the conductivity
$\sigma_{xy}$ observed in the integer QH effect \cite{JanVFH94}.
However, because of its purely classical assumptions the model is unable
to exactly reproduce the critical properties of the transition found in
experiments, i.e., the divergence of the correlation length at the
transition with an exponent of $\nu\approx 2.3$ \cite{KocHKP91a}.
Instead, it predicts $\nu=4/3$, the value appropriate for classical
percolation.

The CC network model improved the high-field model by introducing
quantum corrections \cite{ChaC88}, namely tunneling and interference.
Tunneling occurs, in a semi-classical view, when electron orbits come so
close to each other that the cyclotron orbits overlap. From Fig.\
\ref{fig-qhe-SPnetwork} one can conclude that this happens at the SPs,
which now act as quantum scatterers described by a unitary
scattering matrix $S$\\
\begin{equation}
\left(
\begin{array}{c}
O\\
O^*
\end{array}
\right) =S \left(
\begin{array}{c}
I\\
I^*
\end{array}
\right) = \left(
\begin{array}{cc}
t&r\\
-r&t
\end{array}
\right) \left(
\begin{array}{c}
I\\
I^*
\end{array}
\right)
\end{equation}
which connects two incoming with two outgoing channels. Assuming a
symmetric potential at the SP the scattering rates are given by a pair
of a complex transmission and a reflection coefficient $t$ and $r$,
respectively.  From the required unitarity of $S$ it follows that
$|t|^2+|r|^2=1$.  Obviously, $t$ and $r$ depend on the potential energy
of the SP. It was shown \cite{GalR97} that $t$ and $r$ can be
parametrized by
\begin{equation}
\label{eq-qhe-tz} t=\left(\frac{1}{e^z+1}\right)^\frac{1}{2} \quad
{\rm and} \quad r=\left(\frac{1}{e^{-z}+1}\right)^\frac{1}{2}
\end{equation}
where $z$ corresponds to a dimensionless energy difference between SP
potential and electron energy $\varepsilon$.  Without restricting the
generality $\langle V \rangle=0$ is assumed in the following.  In case
of $\varepsilon=0$ the value of $z$ then coincides with the
dimensionless SP height. Similar to bond percolation now a network of
SPs can be constructed.  The SPs are mapped onto nodes and the
equipotentials correspond to links.
While moving along an equipotential an electron accumulates a
random phase $\Phi$ which reflects the randomness of the
potential. The corresponding phase factor $e^{i\Phi}$ can be
included in the matrix $S$ \cite{JanMMW98} or taken into account
by additional diagonal matrices on each link \cite{GalR97}.

As in the high-field model this quantum percolation model describes only a
single QH transition with exactly one extended state in the middle of
the Landau band at $\varepsilon=0$ .  The critical properties at the
transition, especially the value of the exponent $\nu\approx2.4 \pm 0.2$
\cite{LeeWK93}, agree with experiments \cite{KocHKP91a,SchVOW00} as well
as with results of other theoretical approaches \cite{HucK90,HuoB92}.
For numerical investigations of the CC model, one constructs a regular
2D lattice out of the SPs.  Then the 2D plane is cut into 1D slices with
the associated scattering matrices transformed into a transfer matrix.
The conductivity may be calculated by transversing perpendicular to the
slices along the sample by transfer matrix multiplications
\cite{LeeWK93} according to the Landauer-B\"{u}ttiker approach
\cite{ButILP85}.  The spatial extension of the 2D plane is limited by
the computational effort although an additional disorder averaging over
many samples is not necessary for quasi-1D samples \cite{LeeWK93}.

The CC model is a strong-magnetic-field (chiral) limit of a general
network model, first introduced by Shapiro \cite{Sha82} and later
utilized for the study of localization-delocalization transitions within
different universality
classes\cite{ChaRKH00,FreJM98,FreJM99,Jan98,KagHAC99,MerJH98}. In
addition to the QH transition, the CC model applies to a much broader
class of critical phenomena since the correspondence between the CC
model and thermodynamic, field-theory and Dirac-fermions models
\cite{GruRS97,HoC96,Kim96,KonM97,Lee94,LudFSG94,MarT99,Zir94,Zir97} was
demonstrated.

\subsection{Derivation of the basic RG equation for transmission amplitudes}
\label{sec-rgapproach-derivation}

\begin{figure}[tbh]
 \centerline{\includegraphics[width=0.6\columnwidth]{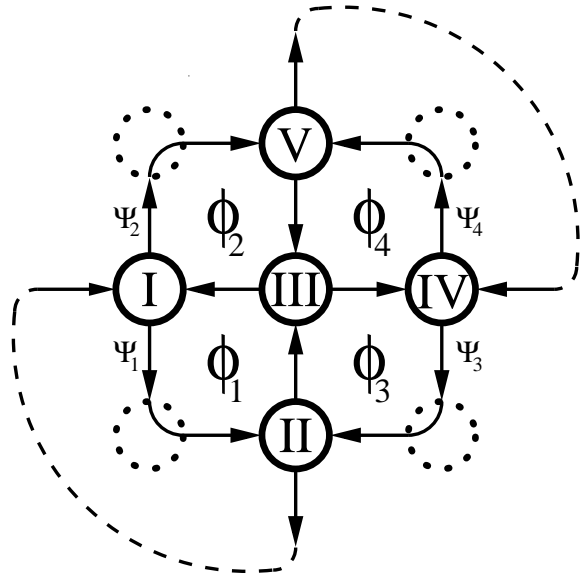}}
\caption{\label{fig-rgstruct}
  CC network on a square lattice consisting of nodes (circles) and links
  (arrows). The RG unit combines five nodes (full circles) by neglecting
  some connectivity (dashed circles).  $\Phi_1, \ldots, \Phi_4$ are the
  phases acquired by an electron along the loops as indicated by the
  arrows. $\Psi_1, \ldots, \Psi_4$ represent wave function amplitudes,
  and the thin dashed lines illustrate the boundary conditions used for
  the computation of level statistics.  }
\end{figure}

The real-space RG approach \cite{AroJS97,GalR97} can be applied to the
CC network analogously to the case of 2D bond percolation
\cite{Ber78,ReyKS77,StaA92}. An RG unit is constructed containing
several SPs from a CC network. For these SPs the RG transformation has
to relate their $S$ matrices with the $S$ matrix of the super-SP. The RG
unit used here is extracted from a CC network on a regular 2D square
lattice.  The super-SP consists of five original SPs connected according
to Fig.\ \ref{fig-rgstruct}.  Circles correspond to SPs and lines to
links in the network. Using this intuitive picture one can identify the
loss of connectivity in comparison with the original CC network, namely,
the four edge nodes within a $3\times 3$ SP pattern are fully neglected
as are their outer bonds. Thus the super-SP has the same number of
incoming and outgoing channels as an original SP.  In analogy to bond
percolation the size of the RG unit in terms of lattice spacings equals
$2$ \cite{CaiRSR01}.

Between the SPs of the RG unit an electron travels along equipotential
lines, and accumulates a certain Aharonov-Bohm phase as in the original
network.  Different phases are uncorrelated, which reflects the
randomness of the original potential landscape.  As shown in section
\ref{sec-rgapproach-ccmodel} each SP is described by an $S$ matrix which
contributes two equations relating the wave-function amplitudes of
incoming $I_i,I_i^*$ and outgoing $O_i,O_i^*$ channels. All amplitudes
$I_i,I_i^*$ besides the external $I_1$ and $I_4^*$ can then be expressed
by $O_i,O_i^*$ using the phases, e.g. $I_5=e^{i \Phi_{15}} O_1$, where
$\Phi_{15}$ is the phase shift along the link between SPs $I$ and $V$.
The resulting ten modified scattering equations form a linear system
which has to solved in order to obtain the transmission properties of
the corresponding super-SP:
\begin{equation}
\label{eq-rgtest-LGS} A{\bf x}={\bf b}
\end{equation}
with
\begin{equation}
\begin{array}{cccc}
A= & \left(
\begin{array}{c}
1\\
0\\
0\\
0\\
0\\
0\\
0\\
0\\
-t_5 e^{i \Phi_{15}}\\
-r_5 e^{i \Phi_{15}}
\end{array}
\right. &
\begin{array}{cccc}
0& 0& 0& 0\\
1& 0& 0& 0\\
-t_2 e^{i \Phi_{12}}& 1& 0& 0\\
-r_2 e^{i \Phi_{12}}& 0& 1& 0\\
0& -r_3 e^{i \Phi_{23}}& 0& 1\\
0& t_3 e^{i \Phi_{23}}& 0& 0\\
0& 0& 0& t_4 e^{i \Phi_{34}}\\
0& 0& 0& -r_4 e^{i \Phi_{34}}\\
0& 0& 0& 0\\
0& 0& 0& 0
\end{array}
& \Rightarrow
\\
\multicolumn{4}{c}{}\\
& \Rightarrow &
\begin{array}{cccc}
 -r_1 e^{i \Phi_{31}}& 0& 0& 0\\
 t_1 e^{i \Phi_{31}}& 0& 0& 0\\
0& 0& -r_2 e^{i \Phi_{42}}& 0\\
 0& 0& t_2 e^{i \Phi_{42}}& 0\\
 0& 0& 0& 0\\
 1& 0& 0& 0\\
 0& 1& 0& 0\\
 0& 0& 1& 0\\
 0& -r_5 e^{i \Phi_{45}}& 0& 1\\
 0& t_5 e^{i \Phi_{45}}& 0& 0
\end{array}
& \left.
\begin{array}{c}
0\\
0\\
0\\
0\\
-t_3 e^{i \Phi_{53}}\\
-r_3 e^{i \Phi_{53}}\\
0\\
0\\
0\\
1
\end{array}
\right),
\end{array}
\end{equation}
\begin{equation}
{\bf
x}=\left(O_1,O_1^*,O_2,O_2^*,O_3,O_3^*,O_4,O_4^*,O_5,O_5^*\right)^T
\end{equation}
and
\begin{equation}
{\bf b}=\left(t_1 I_1, r_1 I_1, 0, 0, 0, 0, r_4 I_4^*, t_4 I_4^*,
0, 0 \right)^T .
\end{equation}
%
Note that the amplitudes on the external links coincide with the
amplitudes of the super-SP as $I_1=I',I_4^*=I'^*,O_5=O'$ and
$O_2^*=O'^*$.  Setting the incoming links of the super-SP according to
$I'=1,I'^*=0$ one can deduce the transmission coefficient $t'$ of the
super-SP, since $O'=t' I'= t'1=t'$.  For the transmission coefficient of
the super-SP this method yields the following expression\cite{GalR97}:
\begin{equation}
  \label{eq-rg-qhrg}
  t'= \left | \frac{
  t_1 t_5 (r_2 r_3 r_4 e^{i \Phi_3} - 1) +
  t_2 t_4 e^{i  (\Phi_1+\Phi_4)} (r_1 r_3 r_5 e^{-i \Phi_2} - 1) +
  t_3 (t_2 t_5 e^{i \Phi_1} + t_1 t_4 e^{i \Phi_4})
} {
  (r_3 - r_2 r_4 e^{i \Phi_3}) (r_3 - r_1 r_5 e^{i \Phi_2}) +
  (t_3 - t_4 t_5 e^{i \Phi_4}) (t_3 - t_1 t_2 e^{i \Phi_1})
}\right | \quad .
\end{equation}
Here $\Phi_j$ corresponds to the sum over the three phases forming a
closed loop within the RG unit (see Fig.\ \ref{fig-rgstruct}).  Equation
(\ref{eq-rg-qhrg}) is the RG transformation, which allows one to
generate the distribution $P(t')$ of the transmission coefficients of
super-SPs using the distribution $P(t)$ of the transmission coefficients
of the original SPs. Since the transmission coefficients of the original
SPs depend on the electron energy $\varepsilon$, the fact that
delocalization occurs at $\varepsilon = 0$ implies that a certain
distribution, $P_{\rm c}(t)$ --- with $P_{\rm c}(t^2)$ being symmetric
with respect to $t^2=\frac{1}{{2}}$ --- is the FP distribution of the RG
transformation (\ref{eq-rg-qhrg}).

A systematic improvement of the RG structure by inclusion of more than
five SPs into the basic RG unit \cite{JanMMW98,JanMW98,WeyJ98} leads to
similar results.
In contrast, using a smaller RG unit \cite{ZulS01} does not describe the
critical properties of the QH transition with the same accuracy. The
super-SP now consists only of $4$ SPs such that it resembles the $5$SP
unit (see Fig.\ \ref{fig-rgstruct}) used previously but leaves out the
SP in the middle of the structure.  Again the scattering problem can be
formulated as a system of now $8$ equations which is solved analytically
to give
\begin{equation}
\label{eq-rg-qhrg_4SP} t'_{\rm 4 SP}=\left|\frac{  t_1 t_4 (r_2
r_3 e^{i \Phi_3} - 1) + t_2 t_3 e^{i \Phi_2}(r_1 r_4 e^{-i \Phi_1}
- 1)} { (1 -r_2 r_3  e^{i \Phi_3})(1 - r_1 r_4 e^{i \Phi_1}) + t_1
t_2 t_3 t_4 e^{i \Phi_2}} \right| \quad.
\end{equation}
The result can be verified using Eq.\ (\ref{eq-rg-qhrg}) after setting
$t_3=0$ and $r_3=1$, joining the phases $\Phi_1$ and $\Phi_4$ and
renumbering the indices. As we will show in section
\ref{sec-pglsd-critexp} the value of $\nu$ obtained for the $4$SP RG
units is less reliable.  Apparently, the quality of the RG approach
crucially depends on the choice of the RG unit. For the construction of
a properly chosen RG unit two conflicting aspects have to be considered.
(i) With the size of the RG unit also the accuracy of the RG approach
increases since the RG unit can preserve more connectivity of the
original network. (ii) As a consequence of larger RG units the
computational effort for solving the scattering problem rises,
especially in the case where an analytic solution, as Eq.\
(\ref{eq-rg-qhrg}), is not attained. Because of these reasons building
an RG unit is an optimization problem depending mainly on the
computational resources available.

\subsection{Description of the RG approach to the LSD}
\label{sec-lsd-model}
The nearest-neighbor level-spacing distribution (LSD) is one of
the important level statistics --- other are $\Sigma_2$ and
$\Delta_3$ statistics --- which has been used extensively
\cite{CaiRR03} to characterize the universal statistical
properties of complex Hamiltonians in the context of random matrix
theory \cite{Meh91}.
The necessary eigenenergies are usually obtained from the
time-independent Schr\"odinger equation $H\Psi=E_k \Psi$ by
diagonalizing the Hamiltonian $H$ \cite{PreFTV92}. After sorting the
eigenenergies in ascending order the LSD is accumulated from spacings
$s_k=(E_{k+1}-E_{k})/\Delta$, where $E_{k+1}$ and $E_k$ are neighboring
energy levels and $\Delta$ corresponds to the mean level spacing. With
the CC model based on wave propagation through the sample, $H$ is not
accessible directly.  In this work therefore an alternative approach \cite{Fer88} is
used in which the energy levels of a 2D CC network can be
computed from the energy dependence of the so-called network
operator $U(E)$.
$U$ is constructed similar to Eq.\ (\ref{eq-rgtest-LGS}).  However, when
comparing to the calculation of the transmission coefficient $t'$ an
essential difference has to be taken into account. Energy levels are
defined only in a closed system which requires to apply appropriate,
usually periodic, boundary conditions.
The energy dependence of $U(E)$ enters through the energy
dependence of the $t_i(E)$ of the SPs, whereas the energy
dependence of the phases $\Phi_j(E)$ of the links is usually
neglected.  Considering the vector $\Psi$ of wave amplitudes on
the links of the network, $U$ acts similar as a time evolution
operator.  The eigenenergies can now be obtained from the
stationary condition
\begin{equation}
\label{eq-lsd-statU} U(E) \Psi = \Psi .
\end{equation}
Nontrivial solutions exist only for discrete energies $E_k$, which
coincide with the eigenenergies of the system\cite{Fer88}. The
eigenvectors $\Psi_k$ correspond to the eigenstates on the links.
The evaluation of the $E_k$'s according to Eq.\
(\ref{eq-lsd-statU}) is numerically very expensive. For that
reason a simplification was proposed\cite{KleM97}. Instead of
solving the real eigenvalue problem calculating a spectrum of
quasienergies $\omega$ is suggested following from
\begin{equation}
\label{eq-lsd-statUomega} U(E) \Psi_l = e^{i  \omega_l(E)} \Psi_l
.
\end{equation}
For fixed energy $E$ the $\omega_l$ are expected to obey the same
statistics as the real eigenenergies\cite{KleM97}. This approach
makes it perfectly suited for large-size numerical simulations,
e.g. studying $50\times50$ SP networks.

In order to combine the above algorithm with the RG iteration, in
which a rather small unit of SPs is considered, some adjustment is
necessary.  First, one has to ``close'' the RG unit at each RG
step in order to discretize the energy levels. From the possible
variants the closing is chosen as shown in Fig.\
\ref{fig-rgstruct} with dashed lines.
For a given closed RG unit with a fixed set of $t_i$-values at the
nodes, the positions of the energy levels are determined by the energy
dependences $\Phi_j(E)$ of the four phases along the loops. These phases
change by $\sim \pi$ within a very narrow energy interval, inversely
proportional to the sample size. Within this interval the change of the
transmission coefficients is negligibly small. The closed RG unit in
Fig.\ \ref{fig-rgstruct} contains $10$ links and thus it is described by
$10$ amplitudes. Each link is characterized by an individual phase. On
the other hand, it is obvious that the energy levels are determined only
by the phases along the loops.  One way to derive $U$ is to combine the
individual phases into phases $\Phi_j$ connected to the four inner loops
of the unit. The $\Phi_j$ are associated with the corresponding
``boundary'' amplitudes $\Psi_j$ (see Fig.\ \ref{fig-rgstruct}). The
original system of 10 equations, which resembles Eq.\
(\ref{eq-rgtest-LGS}) except for the boundary conditions, can then be
transformed to 4 equations by expressing all amplitudes in terms of the
$\Psi_j$.
The resulting network operator takes the form
\begin{equation}
\label{eq-lsd-system}
\begin{array}{cc}
U=&
\begin{array}{ccc}
\left(
\begin{array}{c}
(r_1 r_2 - t_1 t_2 t_3) e^{-i \Phi_1}\\
-t_1 r_3 r_4 e^{-i \Phi_2}\\
-t_1 t_4 r_3 e^{-i \Phi_4}\\
-(t_2 r_1 + t_1 t_3 r_2) e^{-i \Phi_3}
\end{array}
\right. &
\begin{array}{c}
(t_1 r_2 + t_2 t_3 r_1) e^{-i \Phi_1}\\
r_1 r_3 r_4 e^{-i \Phi_2}\\
t_4 r_1 r_3 e^{-i \Phi_4}\\
-(t_1 t_2 - t_3 r_1 r_2) e^{-i \Phi_3}
\end{array}
& \Rightarrow
\end{array}\\
\multicolumn{1}{c}{}\\
&
\begin{array}{ccc}
\Rightarrow &
\begin{array}{c}
t_2 t_5 r_3 e^{-i \Phi_1}\\
-(t_4 r_5 + t_3 t_5 r_4) e^{-i \Phi_2}\\
(r_4 r_5 - t_3 t_4 t_5) e^{-i \Phi_4}\\
t_5 r_2 r_3 e^{-i \Phi_3}
\end{array}
& \left.
\begin{array}{c}
t_2 r_3 r_5 e^{-i \Phi_1}\\
(t_4 t_5 - t_3 r_4 r_5) e^{-i \Phi_2}\\
-(t_5 r_4 + t_3 t_4 r_5) e^{-i \Phi_4}\\
r_2 r_3 r_5 e^{-i \Phi_3}
\end{array}
\right),
\end{array}
\end{array}
\end{equation}
which can be substituted in Eq.\ (\ref{eq-lsd-statUomega}).  Then the
energy levels $E_k$ of the closed RG unit including phases
$\Phi_j(E)=\Phi_j(E_k)$, are the energies for which one of the four
eigenvalues of the matrix $U$ is equal to one, which corresponds to the
condition $\omega(E_k)=0$.  Thus, the calculation of the energy levels
reduces to a diagonalization of the $4\times 4$ matrix.  It should be
emphasized that the reduced size of $U$ in comparison with the
$10\times10$ matrix resulting from the ``straightforward'' approach
described in the beginning directly follows from considering only the
relevant energy dependence in the four phases of the RG unit.  Therefore
a larger size of $U$ would lead to redundant information for the energy
levels.

\subsection{Computing resistances from the transmission coefficients}

We first establish how to connect the experimentally relevant
conductances and resistances to the transmission coefficients. The
dimensionless conductance is simply $G=|t|^2$. The dimensionless
longitudinal resistance $R_{\rm L}$ can be computed from $G$ via
\begin{equation}
\label{eq-RL}
    R_{\rm L} = \frac{|r|^2}{|t|^2} = \frac{1 - |t|^2}{|t|^2} =
    \frac{1}{G} - 1 \equiv R_{\rm 2t} - 1
\end{equation}
with the dimensionless $2$-terminal resistance $R_{\rm 2t}$.

\begin{figure}[tbh]
\centerline{\includegraphics[height=0.4\columnwidth]{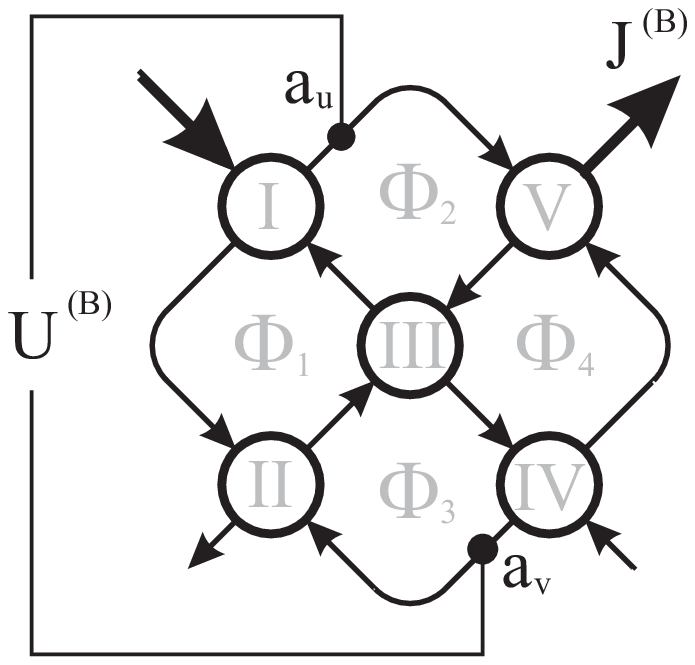}
\quad\includegraphics[height=0.4\columnwidth]{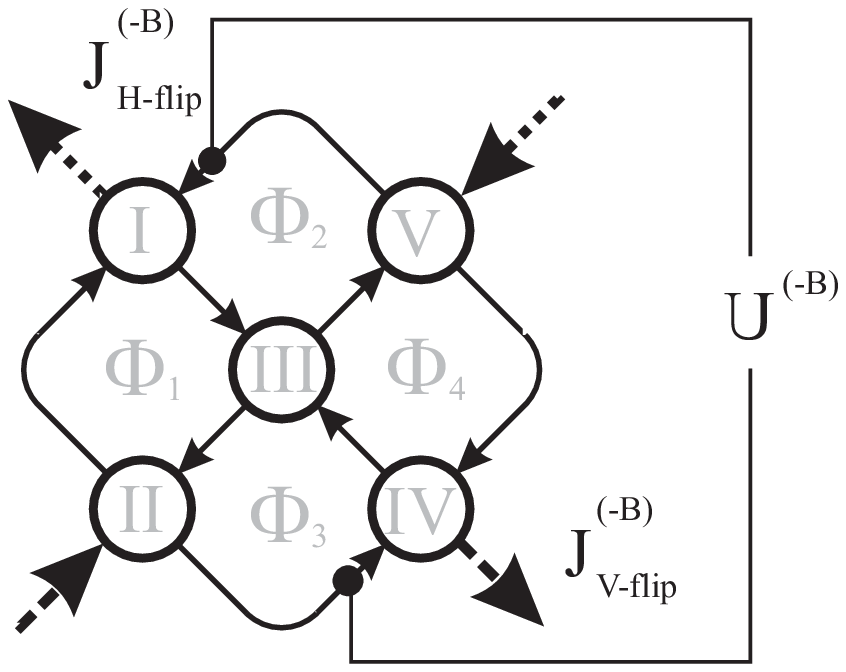}}
\caption{\label{fig-rhall-SP5} To determine the Hall resistance
  $R_{\rm H}$ of the RG unit as illustrated here for the $5$SP case
  the voltage $U$ and the current $J$ are calculated for positive
  magnetic field $B$ (left) and the opposite direction $-B$
  (right). For $-B$ the value of $R_{\rm H}$ can be obtained by
  measuring $J$ either between the upper (H-flip) or lower
  (V-flip) two links, respectively.}
\end{figure}

The computation of $R_{\rm H}$ is less straightforward. We
calculate the ``resistance'' $R=U/J$ defined by the potential
difference $U$ across the RG unit and the current $J$ flowing
through the unit. In Fig.\ \ref{fig-rhall-SP5} this ansatz is
illustrated for the $5$SP RG unit used previously.
Assuming that the current enters the RG unit via one incoming link
($I'$) only and the other incoming link is inactive ($I'^*=0$) the
resulting power transmission coefficient $t'^2=O'^2/I'^2$ of the
RG unit can be associated with $J$. In order to determine $U$ one
considers the quantities $a_u=O_1^2/I'^2$ and $a_v=O_4^{*2}/I'^2$
as chemical potentials measured by weakly coupled voltage probes
at these opposite links of the RG unit \cite{BeeH91}. Thus the
voltage drop is given by $U=a_u-a_v$. Because of the four-terminal
geometry the obtained $R$ contains beside the Hall resistance
$R_{\rm H}$ also a contribution from the longitudinal resistance
$R_{\rm L}$. The separation of the unwanted part $R_{\rm L}$ is
accomplished by employing the antisymmetry of the Hall voltage
$U_{\rm H}$ under the reversal of the direction of the magnetic
field $B$
\begin{equation}
U_{\rm H}=\frac{1}{2} \left[U^{(B)}-U^{(-B)}\right] .
\end{equation}
Considering the $5$SP RG unit again one therefore obtains
\begin{equation}
\label{eq-rhall-rg} R_{\rm H}=\frac{1}{2}
\frac{\left[a_u^{(B)}-a_v^{(B)}\right] -
  \left[a_u^{(-B)}-a_v^{(-B)}\right]}{t'^2} .
\end{equation}
The quantities $a_u^{(B)}$ and $a_v^{(B)}$ are calculated by
solving the system of equations (\ref{eq-rgtest-LGS}) analogously
to the determination of $t'$ in section
\ref{sec-rgapproach-ccmodel}. One finds
\begin{equation}
  a_u^{(B)}=
\left| \frac{  t_1 (1 -  r_2 r_3 r_4 e^{i \Phi_3}-  t_3 t_4 t_5
e^{i \Phi_4}) - t_2 e^{i \Phi_1} (t_3 -  t_4 t_5 e^{i \Phi_4})}
{(r_3 -  r_2 r_4 e^{i \Phi_3}) (r_3 -  r_1 r_5 e^{i \Phi_2}) +
(t_3 - t_4 t_5 e^{i \Phi_4})(t_3 -  t_1 t_2 e^{i \Phi_1}) }
\right|^2
\end{equation}
and
\begin{equation}
a_v^{(B)}= \left| \frac{r_4 ( r_1 r_3 t_2 e^{i \Phi_1} -  r_5 t_2
e^{i (\Phi_1 + \Phi_2)} +  r_5 t_1 t_3 e^{i \Phi_2})} {(r_3 -  r_2
r_4 e^{i \Phi_3}) (r_3 -  r_1 r_5 e^{i \Phi_2}) + (t_3 - t_4 t_5
e^{i \Phi_4})(t_3 -  t_1 t_2 e^{i \Phi_1}) } \right|^2 .
\end{equation}
Under the reversed field $-B$ the electrons travel along the same
equipotentials but in the opposite direction. In the corresponding RG
unit the links therefore only change their direction, as shown in Fig.\
\ref{fig-rhall-SP5}.
A full rederivation of the RG equation for $-B$ can
be omitted if the structure of the RG unit is taken into account.
The result with identical directions of the current is obtained by
flipping the unit vertically (V-flip) while the other case
corresponds to a horizontal flip (H-flip). The comparison to the
original RG unit then shows how to map the indices of the SPs and
the phases in order to adopt the RG equation to $-B$, which is
demonstrated here for the
H-flip result where $(r_1,t_1)\leftrightarrow (r_5,t_5)$,
$(r_2,t_2)\leftrightarrow (r_4,t_4)$, and
$\Phi_1\leftrightarrow \Phi_4$,
\begin{eqnarray}
a_u^{(-B)}&= &\left| \frac{ t_5 (1 -  r_2 r_3 r_4 e^{i \Phi_3} -  t_1 t_2
t_3 e^{i \Phi_1}) -  t_4  e^{i \Phi_4} (t_3 -  t_1 t_2 e^{i \Phi_1}) }
{(r_3 - r_2 r_4 e^{i \Phi_3}) (r_3 - r_1 r_5 e^{i \Phi_2}) + (t_3 -  t_4
t_5 e^{i \Phi_4}) (t_3 - t_1 t_2 e^{i \Phi_1})}
\right|^2 , \\
a_v^{(-B)}&= &\left| \frac{r_2 ( r_3 r_5 t_4 e^{i \Phi_4} -  r_1 t_4 e^{i
(\Phi_2 + \Phi_4)} +  r_1 t_3 t_5 e^{i \Phi_2})}
{(r_3 - r_2 r_4 e^{i \Phi_3}) (r_3 - r_1 r_5 e^{i \Phi_2}) + (t_3 -  t_4
t_5 e^{i \Phi_4}) (t_3 - t_1 t_2 e^{i \Phi_1})}
\right|^2 .
\end{eqnarray}
Using Eq.\ (\ref{eq-rhall-rg}) one is now able to determine the
distributions of $P(R_{\rm H})$ at the QH transition iteratively
in course of the RG iterations.

\subsection{Iterating the RG structures}
\label{sec-rgapproach-numerics}
In order to find the FP distribution $P_{\rm c}(t)$,
the RG is started from a certain initial distribution of
transmission coefficients, $P_0(t)$.  The distribution is
typically discretized in at least $1000$ bins, such that the bin
width is typically $0.001$ for the interval $t\in [0, 1]$. From
$P_0(t)$, the $t_i$, $i=1,\ldots,5$, are obtained and substituted
into the RG transformation (\ref{eq-rg-qhrg}).  The phases
$\Phi_j$, $j= 1,\ldots,4$ are chosen randomly from the interval
$\Phi_j \in [0, 2\pi)$.  In this way at least $10^{7}$
super-transmission coefficients $t'$ are calculated. In order to
decrease statistical fluctuations the obtained histogram $P_1(t')$
is then smoothed using a Savitzky-Golay filter \cite{PreFTV92}. At
the next step the procedure is repeated using $P_1$ as an initial
distribution.  The convergence of the iteration process is assumed
when the mean-square deviation $\int{dt [
  P_n(t)-P_{n-1}(t) ]^2}$ of the distribution $P_n$ and its
predecessor $P_{n-1}$ deviate by less than $10^{-4}$.

The actual initial distributions, $P_0(t)$, were chosen in such a
way that the corresponding conductance distributions, $P_0(G)$,
were either uniform or parabolic, or identical to the FP
distribution found semianalytically \cite{GalR97}.  All these
distributions are symmetric with respect to $G=0.5$. One can
observe that, regardless of the choice of the initial
distribution, after $5$--$10$ steps the RG procedure converges to
the {\em same} FP distribution which remains unchanged for another
$4$--$6$ RG steps.  Small deviations from the symmetry with
respect to $G=0.5$ finally accumulate due to numerical
instabilities in the RG procedure, so that typically after
$15$--$20$ iterations the distribution becomes unstable and flows
toward one of the classical FPs $P(G)=\delta(G)$ or
$P(G)=\delta(G-1)$.  Therefore the symmetry of the $P_0(G)$ with
respect to $G=0.5$ is an important requirement in order to
converge to the quantum FP at all.  Note that the FP distribution
can be stabilized by forcing $P_n(G)$ to be symmetric with respect
to $G=0.5$ in the course of the RG procedure.

Since the dimensionless SP height $z_i$ and the transmission
coefficient $t_i$ at $\varepsilon=0$ are related by Eq.\
(\ref{eq-qhe-tz}), transformation (\ref{eq-rg-qhrg}) also
determines the height of a super-SP by the heights of the five
constituting SPs. Correspondingly, the distribution $P(G)$
determines also a distribution $Q(z)$ of the SP heights via $Q(z)
= P(G) |dG/dz| = \frac{1}{4}\cosh^{-2}(z/2)
P\left[(e^z+1)^{-1}\right]$.
This represents a convenient parameterization of the conductance
distribution, particularly if small and large $t$ values become
important. In this case, it is numerically better to perform the
RG approach using the $Q(z)$ distribution.
We typically discretize the distribution $Q(z)$ in at least $6000$ bins
such that the bin width is $0.01$. Since $z \in ]-\infty, \infty[$, we
have to include lower and upper cut-off SP heights such that $z \in
[z_{\rm low}, z_{\rm up}]$. From $Q_0(z)$, we obtain $z_i$,
$i=1,\ldots,5$, compute the associated $t_i$ and substitute them into
the RG transformation (\ref{eq-rg-qhrg}).  At the next step we repeat
the procedure using the new $Q_1$ as an initial distribution.  We check
that the values of $z_{\rm low,up}$ do not influence our results.

\section{Results of the RG procedure at the QH transition}
\label{sec-pglsd}

\subsection{Fixed point distributions and the critical exponent}
\label{sec-pglsd-fpdist-critexp}

\subsubsection{The fixed point distributions $P_{\rm c}(G)$,
$Q_{\rm c}(z)$} \label{sec-pglsd-pgqz}

\begin{figure}[tbh]
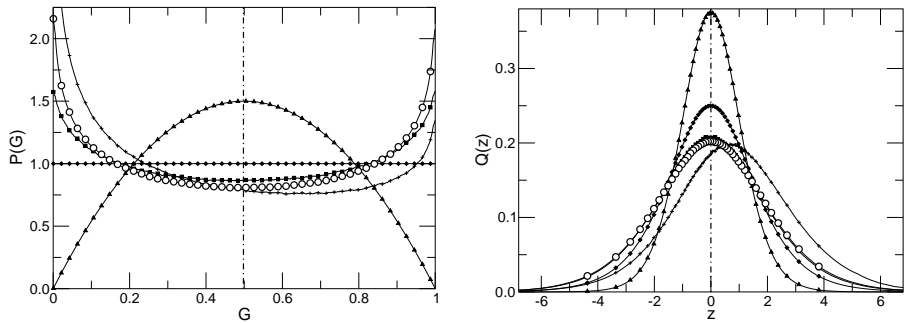

\centerline{\includegraphics[width=0.45\columnwidth]{P0.eps}
\quad\includegraphics[width=0.45\columnwidth]{Q0.eps}}
\caption{\label{fig-rg-PQ}
  Left: $P(G)$ (thin lines) as function of conductance $G$ at a QH
  plateau-to-plateau transition.  Symbols mark every $20$th data point
  for different initial distributions
  ($\blacksquare$,$\blacklozenge$,$\blacktriangle$), the FP distribution
  ($\bigcirc$) and a distribution for RG step $n=16$ ($+$). The vertical
  dashed line indicates the average of the FP distribution.  Right:
  Corresponding plots for the distribution $Q(z)$ of SP heights.}
\end{figure}

The distribution $P_{\rm c}(G)$ of the dimensionless conductance
$G$ can be obtained from $P(t)$ so that
\begin{equation}
P_{\rm c}(G)\equiv \frac{1}{2t}P_{\rm c}(t) \quad.
\end{equation}
Figure \ref{fig-rg-PQ} illustrates the RG evolution of $P(G)$ and
$Q(z)$. In order to reduce statistical fluctuations the FP
distribution is averaged over nine results obtained from three
different $P_0(G)$'s.  The FP distribution $P_{\rm c}(G)$ exhibits
a flat minimum around $G=0.5$, and sharp peaks close to $G=0$ and
$G=1$. It is symmetric with respect to $G\approx 0.5$ with
$\langle G \rangle = 0.498\pm0.004$, where the error is the
standard error of the mean of the obtained FP distribution.
Consequently, the FP distribution $Q_c(z)$ is symmetric with
respect to $z=0$, which corresponds to the center of the Landau
band.  The shape of $Q_c(z)$ is close to Gaussian.
We note that the present results agree with
\cite{JanMMW98,JanMW98,WeyJ98} where a similar RG treatment of the CC
model was carried out.  Our numerical data have a higher resolution, and
show significantly less statistical noise because of the faster
computation using the analytical solution (\ref{eq-rg-qhrg}) of the
RG\cite{GalR97}.

\subsubsection{The fixed point distributions $P(R_{\rm H})$,
$P(R_{\rm L})$} \label{sec-pglsd-prhprl}

In order to accurately construct the tails of the resistance
distributions $P(R_{\rm L})$, $P(R_{\rm H})$, we employ the
numerical strategy of section \ref{sec-rgapproach-numerics} and
discretize $Q(z)$ with $z_{\rm low}= -20$, $z_{\rm up}= 40$ for
perturbations towards positive $z$ and vice versa for negative
$z$. This corresponds to transmission amplitudes $t(z_{\rm
low})\approx 10^{-5}$, $t(z_{\rm
  up})\approx 1-10^{-9}$ for positive perturbations and vice versa for
negative perturbations. Next, we construct via the RG procedure the
sequence of distribution $Q_n(z)$, $n=1, 2, 3, \ldots$, calculating at
least $10^{8}$ super-transmission coefficients $t'$ and associated $z'$.
We assume that the iteration process has converged when the mean square
deviation $\int{dz \left[Q_n(z)-Q_{n-1}(z) \right]^2}$ of the
distribution $Q_n$ and its predecessor $Q_{n-1}$ deviate by less than
$10^{-4}$. We check that the values of $z_{\rm low,up}$ do not influence
our results and that all the previous results at the QH transition as in
Refs.\ \cite{CaiRSR01} are reproduced. The full width at half maximum of
the FP distribution $Q_{\rm c}(z)$ is about $5$ \cite{CaiRSR01}.

Using Eq.\ (\ref{eq-rhall-rg}) we now obtain besides the FP
distribution $Q_{\rm c}(z)$ also the FP distribution $P_{\rm
c}(R_{\rm H})$. Like $z$ the value of $R_{\rm H}$ is not bound to
a fixed interval. To perform a discretization requires to set
lower and upper bounds appropriately. The resulting histogram is
limited to $[-100,100]$ containing $40000$ bins.  Since $P_{\rm
c}(t)$ is already known from section \ref{sec-pglsd-pgqz} one can
speed up the determination by using $Q_{\rm
  c}(z)$ as initial $Q_0(z)$ and therefore obtain $P_{\rm c}(R_{\rm H})$
already after the first iteration.

\begin{figure}[tbh]
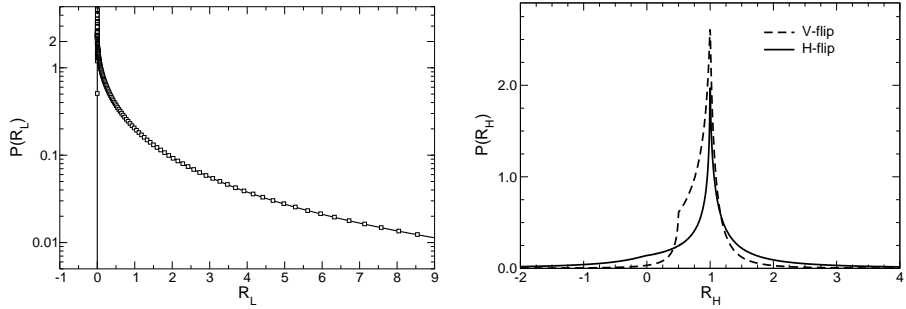

  \centerline{\includegraphics[width=0.45\columnwidth]{fig-P_RL-FP.eps}
\quad
\includegraphics[width=0.45\columnwidth]{fig-P_RH-FP.eps}}
\caption{\label{fig-rhall-FP-5SP} Left: Distribution of the
longitudinal resistance $R_{\rm L}$. The $\Box$ symbols indicate
the FP distribution, the solid line is a fit to a log-normal
distribution for positive $R_{\rm L}$. Only every 5th data point
is shown.
Right: Distribution of the Hall resistance $R_{\rm H}$ at the FP
as obtained from the V-flip (dashed line) and the H-flip method
(solid line). Note the kink at $R_{\rm H}=1$ for the V-flip
structure.}
\end{figure}

In Figure \ref{fig-rhall-FP-5SP} the resulting distributions $P_{\rm
  c}(R_{\rm L})$ and $P_{\rm c}(R_{\rm H})$ have been plotted. Both
distributions are strongly non-Gaussian with long tails. $P_{\rm
  c}(R_{\rm L})$ can be fitted by a log-normal distribution
\cite{PryA99}. We have computed $P_{\rm c}(R_{\rm H})$ using both V- and
H-flip structures. Their graphs in Fig.\ \ref{fig-rhall-FP-5SP} show a
significant difference in the shape of $P_{\rm c}(R_{\rm H})$.  The
H-flip distributions are characterized by a very sharp, nearly symmetric
peak at $R_{\rm H}=1$ which coincides with the value of $R_{\rm H}$ at
the first Landau level.
%
For V-flip an asymmetric distribution is found again with a strong peak
at $R_{\rm H}=1$. Surprisingly, there exists also a kink at $R_{\rm
  H}=0.5$. Such a peak would normally correspond to a second Landau
level, which is however not part of the CC model.
The origin of the ``wrong'' kink must stem from a difference between the
calculation of $R_{\rm H}$ according to H-flip and V-flip method,
respectively. Both methods differ only in the $(-B)$ part of $R_{\rm
  H}$. As illustrated in Fig.\ \ref{fig-rhall-SP5}, the H-flip method
measures $J$ between the same SPs $I$ and $V$ as for $(+B)$. In
contrast, the $J$ for V-flip is obtained between SPs $II$ and $IV$.
While one can expect the same result for H-flip and V-flip when
considering only the $(-B)$ contribution separately, the final V-flip
result for $R_{\rm H}$ according to Eq.\ (\ref{eq-rhall-rg}) thus
corresponds to a measurement in a {\em different} sample with different
random potential. We therefore only use the H-flip results in the
following.

\subsubsection{Critical exponent} \label{sec-pglsd-critexp}

The language of the SP heights provides a natural way to extract
the critical exponent $\nu$.  Suppose that the RG procedure starts
with an initial distribution $Q_0(z)=Q_c(z-z_0)$ that is shifted
from the critical distribution $Q_c(z)$ by a small $z_0\propto
\varepsilon$. The meaning of $z_0$ is an additional electron
energy measured from the center of the Landau band.  The fact that
the QH transition is infinitely sharp at $z_0 = 0$ implies that
for any $z_0\ne 0$, the RG procedure drives the initial
distribution $Q(z-z_0)$ away from the FP. Since $z_0\ll 1$, the
first RG step would yield $Q_c(z-\tau z_0)$ with some number
$\tau$ independent of $z_0$. At the $n$th step the center of the
distribution will be shifted by $z_{{\rm max},n}=\tau^n z_0$,
while the sample size will be magnified by $2^n$. After a certain
number of steps, say $n_L$, the shift will grow to
\begin{equation}
\label{eq-rg-nudev1} z_{{\rm max},n_L}=\tau^{n_L} z_0\sim 1
\end{equation}
where a typical SP is no longer transmittable.  Then the
localization length $\xi$ can be identified with the system size
$2^{n_L}a$ where $a$ is the lattice constant of the original RG
unit. Using this relation one can rewrite $\tau^{n_L}$ in Eq.\
(\ref{eq-rg-nudev1}) as a power of $(\xi/a)$
\begin{equation}
(\xi/a)^{(\ln \tau/ \ln 2)}  z_0\sim 1
\end{equation}
from which  follows that $\xi$ diverges as
\begin{equation}
\label{eq-rg-xi-divergence} \xi \sim a z_0^{-(\ln 2/\ln \tau)}=a
z_0^{-\nu}
\end{equation}
with $\nu=\ln 2/\ln\tau$.  When the RG procedure is carried out
numerically, one should check that $z_0$ is small enough so that
$z_{{\rm max},n}\propto z_0$ for large enough $n$. Consequently,
the working formula for the critical exponent can be presented as
\begin{equation}
\label{eq-rg-nu}
 \nu = \frac{\ln 2^n}{\ln \left(\frac{z_{{\rm max},n}}{z_0}\right)}
\end{equation}
which should be independent of $n$ for large $n$.

\begin{figure}[tbh]
\centerline{\includegraphics[width=0.7\columnwidth]{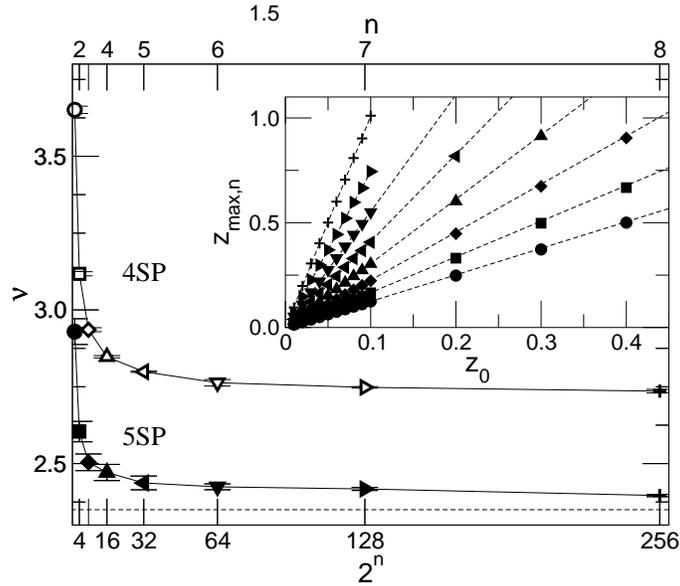}}
\caption{\label{fig-rg-nu-L}
  Critical exponent $\nu$ obtained by the QH-RG approach as function
  of effective linear system size $L= 2^n$ for RG step $n$.
  Solid and open symbols correspond to the 5SP and 4SP RG unit, respectively.
  The error
  bars correspond to the error of linear fits to the data.  The dashed
  line shows $\nu=2.39$. Inset: $\nu$ (here for 5SP) is determined by the dependence
  of the maximum $z_{{\rm max},n}$ of $Q_n(z)$ on a small initial
  shift $z_0$. Symbols indicate the eight RG steps in accordance with
  the main plot. Dashed lines indicate the linear fits.}
\end{figure}

Thus, starting with an initial shift of $Q_c(z)$ by a value $z_0$
results in the further drift of the maximum position, $z_{{\rm
max},n}$, away from $z=0$ after each RG step.  As expected,
$z_{{\rm max},n}$ depends linearly on $z_0$ as shown in Fig.\
\ref{fig-rg-nu-L} (inset) for different $n$ from $1$ to $8$.  The
critical exponent is calculated from the slope according to Eq.\
(\ref{eq-rg-nu}). Its value converges with $n$ to $2.39\pm0.01$.
The error corresponds to a confidence interval of $95\%$ as
obtained from the fit to a linear behavior.
In Fig.\ \ref{fig-rg-nu-L}, we also show results obtained for the
$4$SP structure discussed in section
\ref{sec-rgapproach-derivation}. Obviously, $\nu_{\rm 4SP}$ tends
towards a different value.

\subsubsection{Comparison with other works} \label{sec-pglsd-comp}

By dividing the CC network into units, the RG approach completely
disregards at each RG step the interference of the wave-function
amplitudes between different units. For this reason it is not
clear to what extent this approach captures the main features and
reproduces the quantitative predictions at the QH transition.
Therefore, a comparison of the RG results with the results of
direct simulations of the CC model is crucial. These direct
simulations are usually carried out by employing either the
quasi-1D version \cite{MacK81,ChaC88,LeeWK93} or the 2D version
\cite{FisL81,ChoF97,JovW98,WanJL96} of the transfer-matrix method.
For the critical exponent the values $\nu=2.5\pm0.5$\cite{ChaC88}
and later $\nu=2.4\pm0.2$ \cite{LeeWK93} were obtained.  Note that
the result of the previous section is in excellent agreement with
these values, and is also consistent with the most precise
$\nu=2.35\pm 0.03$\cite{Huc92}. This already indicates the
remarkable accuracy of the RG approach using the $5$SP structure.

In \cite{WanLS98} and \cite{AviBB99} $P_{\rm c}(G)$ was studied by
methods which are not based on the CC model.  Both works reported
a broad distribution $P_{\rm c}(G)$. In
\cite{ChoF97,JovW98,WanJL96} the critical distribution $P_{\rm
  c}(G)$ of the conductance was studied.  $P_{\rm c}(G)$ was found to
be broad, which is in accordance with Fig.\ \ref{fig-rg-PQ}. One
can compare the moments of $P_{\rm c}(G)$ to those calculated in
\cite{WanJL96,ChoF97}. They agree with the present RG
calculations up to the sixth moment. We emphasize that the moments
from \cite{WanJL96} can hardly be distinguished from the moments
of a uniform distribution. This reflects the fact that $P_{\rm
c}(G)$ is practically flat except for the peaks close to $G=0$ and
$G=1$.

\subsection{Energy-level statistics at the QH transition}
\label{sec-pglsd-lsd}

\subsubsection{Choosing an appropriate energy dependence of the
phases} \label{sec-pglsd-lsd-phase}

The crucial ingredient for obtaining the LSD is the choice of the
energy dependence $\Phi_j(E)$. If each loop in Fig.\
\ref{fig-rgstruct} is viewed as a closed equipotential as it is
the case for the first step of the RG procedure\cite{ChaC88}, then
$\Phi_j(E)$ is a true magnetic phase, which changes linearly with
energy with a slope governed by the actual potential profile which
in turn determines the drift velocity. Thus we make the ansatz
\begin{equation}
\label{eq-lsd-PhiE} \Phi_j(E)=\Phi_{0,j}+2\pi\frac{E}{s_j},
\end{equation}
where a random part $\Phi_{0,j}$ is uniformly distributed within
$[0, 2\pi)$, and $2\pi/s_j$ is a random slope. Here the
coefficient $s_j$ acts as an initial level spacing connected to
the loop $j$ of the RG unit by defining a periodicity of the
corresponding phase.
Strictly speaking, the dependence (\ref{eq-lsd-PhiE}) applies only
for the first RG step.  At each step $n>1$, $\Phi_j(E)$ is a
complicated function of $E$ which carries information about all
energy scales at previous steps. However, in the spirit of the RG
approach, one can assume that $\Phi_j(E)$ can still be linearized
within a relevant energy interval.  The conventional RG approach
suggests that different scales in {\em real} space can be
decoupled.  Linearization of Eq.\ (\ref{eq-lsd-PhiE}) implies a
similar decoupling in {\em
  energy} space.  In the case of phases, a ``justification'' of such a
decoupling is that at each RG step, the relevant energy scale,
that is the mean level spacing, reduces by a factor of four.

With $\Phi_j(E)$ given by Eq.\ (\ref{eq-lsd-PhiE}), the statistics
of energy levels determined by the matrix equation
(\ref{eq-lsd-statUomega}) is obtained by averaging over the random
initial phases $\Phi_{0,j}$ and values $t_i$ chosen randomly
according to a distribution $P(t)$.  For every realization the
levels $E_k$ are computed from the solutions $\omega(E_k)=0$ of
Eq.\ (\ref{eq-lsd-statUomega}) as illustrated in Ref.\
\cite{CaiRR03}.
The energy interval is scanned in discrete energy steps with
$\Delta E=\mbox{min}\{s_j\}/250$ which is adapted to each random
realization of the $\Phi_{0,j}$ and takes the periodicity in Eq.\
(\ref{eq-lsd-PhiE}) and its influence on the behavior of
$\omega(E)$ into account.  In particular, each realization yields
three level spacings which are then used to construct a smooth
LSD.  Thus the situation is comparable with estimating the true
RMT ensemble distribution functions from small, say, $2\times 2$
matrices only\cite{Meh91,Wig51}. The outline of the RG procedure
for the LSD is as follows.  The slopes $s_j$ in Eq.\
(\ref{eq-lsd-PhiE}) determine the level spacings at the first
step. They are randomly distributed with a distribution function
$P_0(s)$. Subsequent averaging over many realizations yields the
LSD, $P_1(s)$, at the second step. Then the key element of the RG
procedure, as applied to the level statistics, is using $P_1(s)$
as a {\em
  distribution of slopes} in Eq.\ (\ref{eq-lsd-PhiE}).  This leads to
the next-step LSD and so on.

The approach of this work relies on the ``real'' eigenenergies of
the RG unit.  The simpler computation of the spectrum of
quasienergies adopted in large-scale simulations within the CC
model \cite{KleM97,Met98b} cannot be applied since the energy
dependence of phases $\Phi_j$ in the elements of the matrix is
neglected and only the random contributions, $\Phi_{0,j}$, are
kept. Nevertheless it is instructive to compare the two
procedures. In Ref.\ \cite{CaiRR03} we have calculated the
dependence of the four quasienergies $\omega_k$ on the energy $E$
 with $t_i$ chosen from the critical distribution $P_{\rm c}(t)$. The energy dependence of the phases
$\Phi_j$ was chosen from LSD of the Gaussian unitary ensemble
(GUE) \cite{Meh91,Wig51} according to Eq.\ (\ref{eq-lsd-PhiE}). We
find that the dependences $\omega(E)$ range from remarkably linear
and almost parallel to strongly nonlinear.
%
\subsubsection{The critical LSD at the QH transition}
\label{sec-pglsd-lsd-critical}

\begin{figure}[tbh]
  \centerline{\includegraphics[height=0.38\columnwidth]{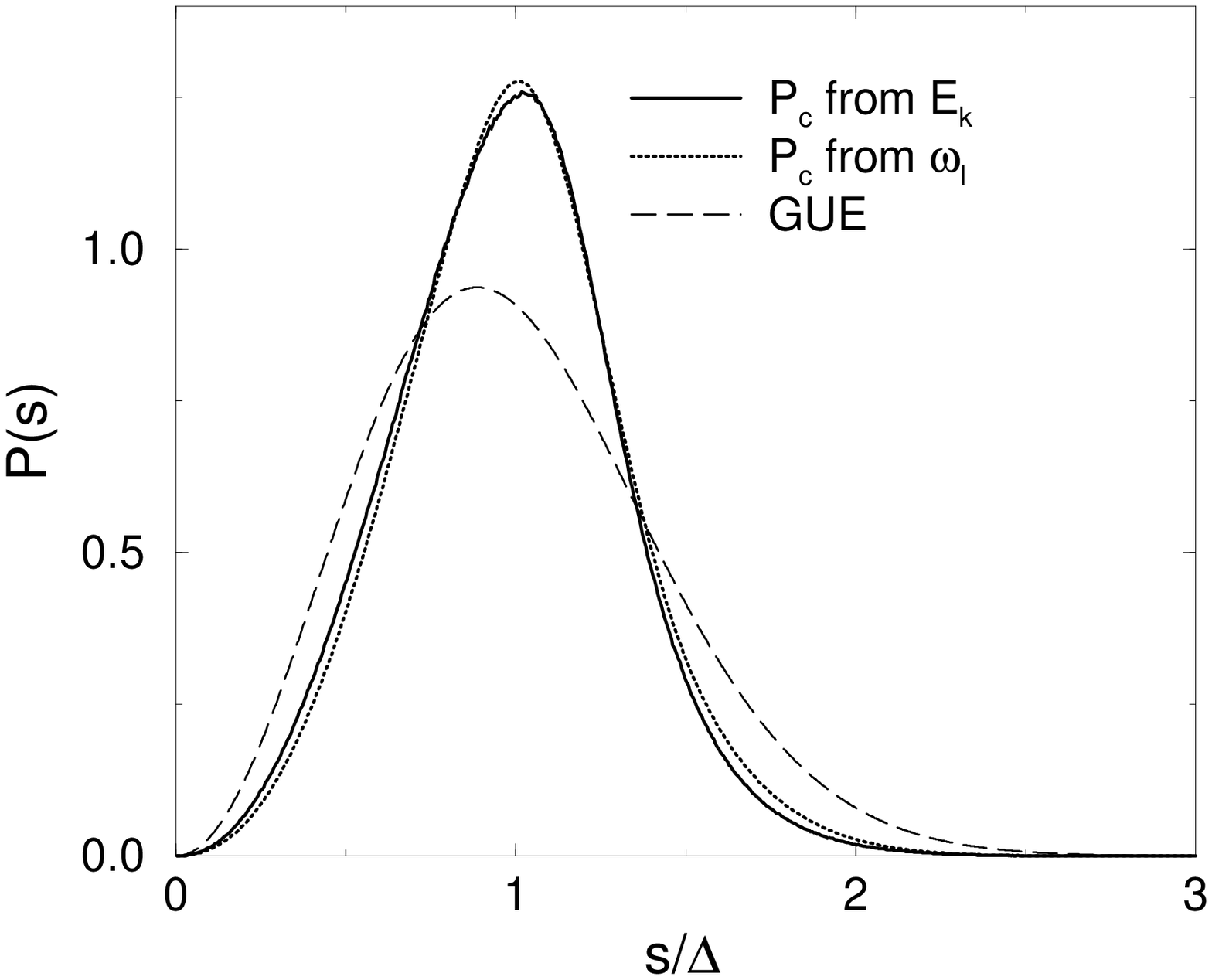}
  \quad \includegraphics[height=0.38\columnwidth]{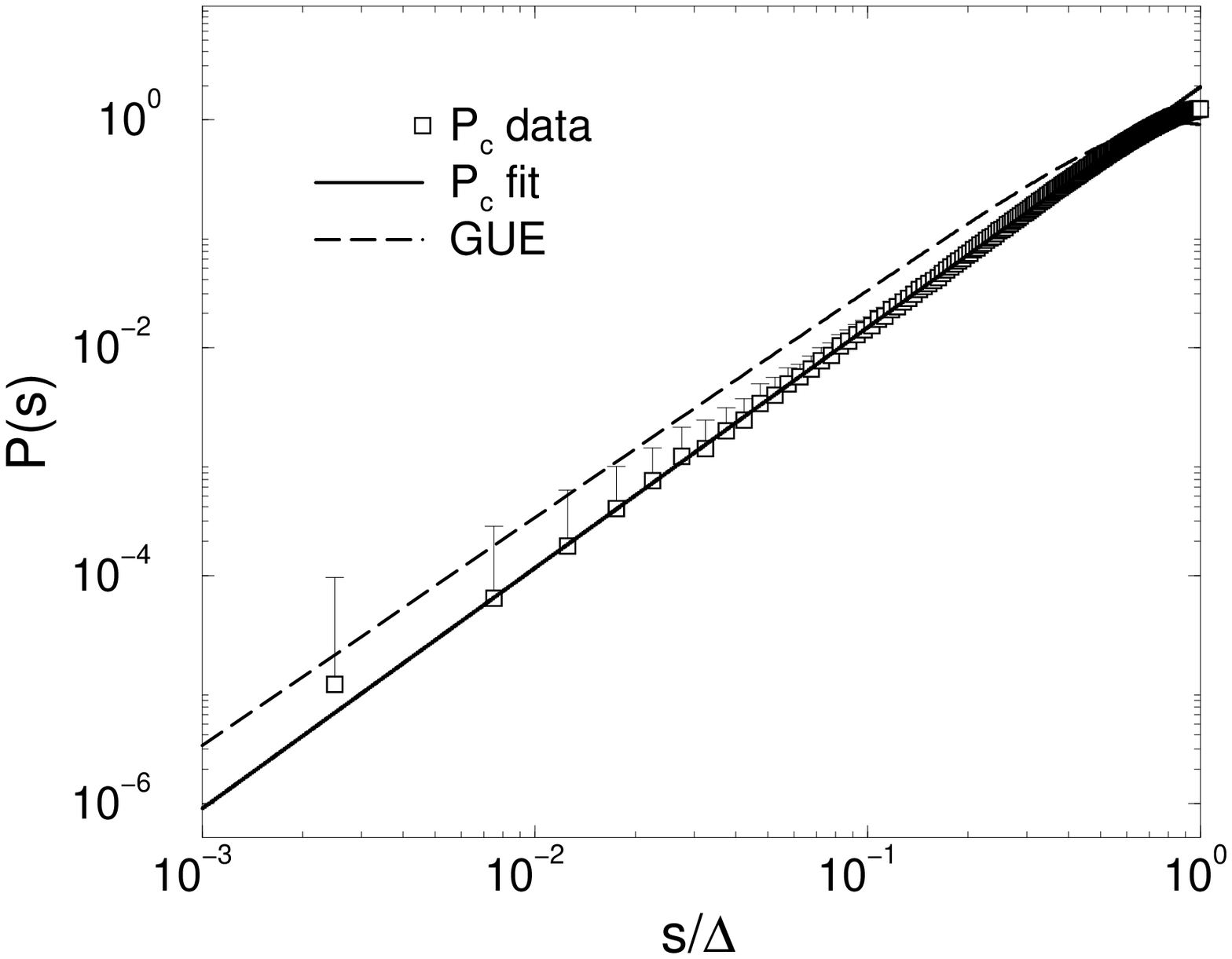}}
\caption{\label{fig-lsd-PsFP}
  Left: FP distributions $P_{\rm c}(s)$ obtained from the spectrum of
  $\omega_l(E=0)$ and from the RG approach using the real eigenenergies
  $E_k$ in comparison to the LSD $P_{\rm GUE}(s)= 32 s^2
  \exp{(-\frac{4}{\pi}s^2)}/\pi^2$ for GUE. As in all other graphs
  $P(s)$ is shown in units of the mean level spacing $\Delta$.
Right: \label{fig-lsd-PsFPsmall}
  $P_{\rm c}(s)$ for small $s$ in agreement with the predicted
  $s^2$ behavior. Due to the log-log plot errors are shown in the
  upper direction only.  }
\end{figure}

Let us first consider the shape of the LSD at the QH transition.
As starting distribution $P_0(s)$ of the RG iteration we choose
the LSD of GUE because previous simulations \cite{BatS96,KleM97}
indicate that the critical LSD is close to GUE. According to
$P_0(s)$, each $s_j$ is selected randomly and $\Phi_j$, $j=
1,\ldots,4$ is set as in Eq.\ (\ref{eq-lsd-PhiE}). For the
transmission coefficients of the SP the FP distribution $P_{\rm
c}(t)$ is used. As known from section \ref{sec-pglsd-pgqz},
$P_n(t)$ drifts away from the FP within several further iterations
due to unavoidable numerical inaccuracies.
In order to stabilize the calculation, the FP distribution $P_{\rm
c}(t)$ is therefore used in every RG step instead of $P_n(t)$.
This trick does not alter the results but speeds up the
convergence of the RG for $P_{\rm c}(s)$ considerably.
By finding solutions $\omega(E_k)=0$ of Eq.\
(\ref{eq-lsd-statUomega}) the new LSD $P_1(s')$ is constructed
from the ``unfolded'' energy level spacings
$s'_k=(E_{k+1}-E_{k})/\Delta$, where $k=1,2,3$, and the mean
spacing $\Delta= (E_4-E_1)/3$.  Due to the
``unfolding''\cite{Haa92} with $\Delta$, the average spacing is
set to one for each sample and in each RG-iteration step spacing
data of $2\times 10^6$ super-SPs are taken into account. The
resulting LSD is discretized in bins with largest width $0.01$.
In the following iteration step the procedure is repeated using
$P_1(s)$ as initial distribution.  Convergence of the iteration
process is assumed when the mean-square deviation of $P_n(s)$
deviates by less than $10^{-4}$ from its predecessor $P_{n-1}(s)$.
The above approach now enables one to determine the critical LSD
$P_{\rm c}(s)$. The RG iteration converges rather quickly after
only $2-3$ RG steps.  The resulting $P_{\rm c}(s)$ is shown in
Fig.\ \ref{fig-lsd-PsFP} together with the LSD for GUE.

Although $P_{\rm c}(s)$ exhibits the expected features, namely,
level repulsion for small $s$ and a long tail at large $s$, the
overall shape of $P_{\rm c}(s)$ differs noticeably from GUE.  In
the previous large-size lattice simulations \cite{BatS96,KleM97}
the obtained critical LSD was much closer to GUE. This fact,
however, does not necessarily imply a lower accuracy of the RG
approach. Indeed, as it was demonstrated recently, the critical
LSD -- although being system-size independent
--- nevertheless depends on the geometry of the samples\cite{PotS98}
and on the specific choice of boundary
conditions\cite{BraMP98,SchP98}.  Sensitivity to the boundary
conditions does not affect the asymptotics of the critical
distribution, but rather manifests itself in the shape of the
``body'' of the LSD. One should note that the boundary conditions
which have been imposed to calculate the energy levels (dashed
lines in Fig.\ \ref{fig-rgstruct}) are {\em non-periodic} in
contrast to usual large-size lattice simulations\cite{KleM97}.

As mentioned above there is also the possibility to assess the
critical LSD via the distribution of {\em
  quasienergies}. In Fig.\ \ref{fig-lsd-PsFP} the result of this
procedure is shown.  It appears that the resulting distribution is
almost {\em identical} to $P_{\rm c}(s)$.  This observation is
highly non-trivial, since there is no simple relation between the
energies and quasienergies \cite{CaiRR03}. Moreover, choosing
another functional form for $\Phi(E)$ instead of the linear
$E$-dependence (\ref{eq-lsd-PhiE}), the RG procedure yields an LSD
which is markedly different (within the ``body'') from $P_{\rm
c}(s)$ \cite{CaiRR03}. Using the quasienergies instead of real
energies, as in \cite{KleM97}, and linearizing the energy
dependence of phases [as in Eq.\ (\ref{eq-lsd-PhiE})] is not
rigorous. However, the coincidence of the results for the two
procedures supports the validity of the approach.

\subsubsection{Small-$s$  and large-$s$ behavior}
\label{sec-pglsd-lsd-smalllarge}

The general shape of the critical LSD is not universal. However,
the small-$s$ behavior of $P_{\rm c}(s)$ must be the same as for
GUE, namely $P_{\rm
  c}(s) \propto s^2$. This is because delocalization at the QH
transition implies level repulsion \cite{FyoM97,ShkSSL93}. Earlier
large-scale simulations of the critical LSD
\cite{BatS96,BatSK98,BatSZK96,FeiAB95,KawOSO96,KleM97,Met98b,Met99,MetV98,OhtO95,OnoOK96}
satisfy this general requirement.  The same holds also for the
result of this work, as can be seen in Fig.\
\ref{fig-lsd-PsFPsmall}. The given error bars of the numerical
data are standard deviations computed from a statistical average
of $100$ FP distributions each obtained for different random sets
of $t_i$'s and $\Phi_j$'s within the RG unit. In general, within
the RG approach, the $s^2$-asymptotics of $P(s)$ is most natural.
This is because the levels are found from diagonalization of the
$4\times 4$ unitary matrix (\ref{eq-lsd-system}) with absolute
values of elements widely distributed between $0$ and $1$.

The right form of the large-$s$ tail of $P(s)$ is Poissonian with
$P_{\rm c}(s)\propto \exp(-bs)$
\cite{ShkSSL93,BatSZK96,BatS96,ZhaK97}. It appears that the
accuracy of the present RG approach is insufficient to discern
this non-trivial feature of the critical LSD, since we have a good
accuracy only for $s/\Delta \lesssim 2.5$ \cite{CaiRR03}.

\subsubsection{Finite-size scaling at the QH transition}
\label{sec-pglsd-els-fss}

In order to extract $\nu$ from the LSD the one-parameter-scaling
hypothesis \cite{AbrALR79} is employed. The approach describes the
rescaling of a quantity $\alpha(N; \{z_i\})$ --- depending on
(external) system parameters $\{z_i\}$ and the system size $N$ ---
onto a single curve by using a scaling function $f$
 \begin{equation}
 \label{eq-lsd-scal}
 \alpha\left(N;\{z_i\}\right)=f\left(\frac{N}{\xi_\infty(\{z_i\})}\right) .
 \end{equation}
We now use the scaling assumption to extrapolate $f$ to
$N\rightarrow\infty$ from the finite-size results of the
computations.  The knowledge about $f$ and $\xi_\infty$ then
allows to derive the value of $\nu$ similarly to
(\ref{eq-rg-xi-divergence}). In order to define a suitable control
parameter $z_0$ in the transition region, again the natural
parameterization (\ref{eq-qhe-tz}) of the transmission
coefficients is used.
Because the QH transition occurs exactly at $z_0 = 0$ any initial
distribution $Q(z-z_0)$ with $z_0\ne 0$ will evolve during the RG
procedure toward an insulator, either with complete transmission
of the network nodes (for $z_0>0$) or with complete reflection of
the nodes (for $z_0<0$).

\begin{figure}[tbh]
  \centerline{\includegraphics[width=0.7\columnwidth]{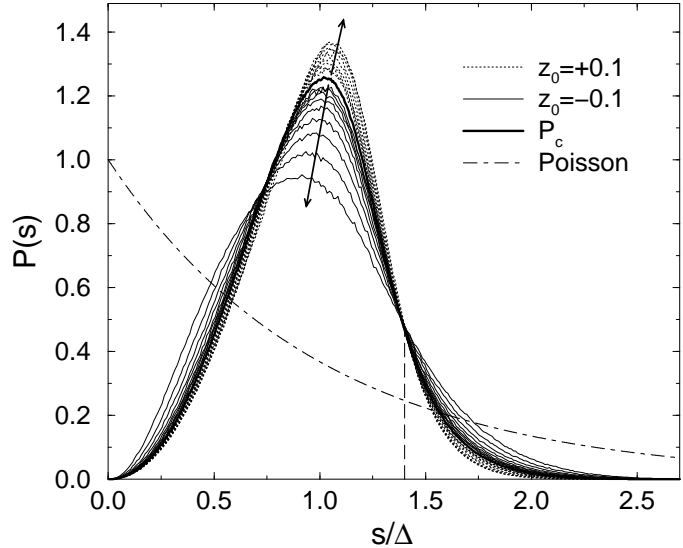}}
\caption{\label{fig-lsd-LSDshift}
  RG of the LSD used for the computation of $\nu$. The dotted lines
  corresponds to the first $9$ RG iterations with an initial
  distribution $P_0$ shifted to the metallic regime ($z_0=0.1$) while
  the full thin lines represent results for a shift toward
  localization ($z_0=-0.1$). Within the RG procedure the LSD moves
  away from the FP as indicated by the arrows. At $s/\Delta\approx
  1.4$ the curves cross at the same point which is used
  when deriving a scaling quantity from the LSD.  }
\end{figure}

In principle, one is free to choose for the finite-size scaling
(FSS) analysis any characteristic quantity $\alpha(N;z_0)$
constructed from the LSD which has a systematic dependence on
system size $N$ for $z_0\ne0$ while being constant at the
transition $z_0=0$.  Out of the large number of possible
choices\cite{BatS96,HofS94b,ShkSSL93,ZhaK95b,ZhaK95c,ZhaK97} a
restriction is made to quantities that are defined mainly by the
small-$s$
 behavior which is accurately described by the RG approach.  The
quantities are obtained by integration of the LSD and have already
been successfully used in \cite{HofS93,HofS94b,ZhaK95b}, namely
\begin{equation}
  \alpha_{\rm P} = \int^{s_0}_0 P(s) ds,
\quad
  \alpha_{\rm I} = \frac{1}{s_0}\int^{s_0}_0 I(s)ds ,
\quad
    \alpha_{\rm S} = \int^{\infty}_0 s^2 P(s) ds
\end{equation}
with $I(s)= \int^{s}_0 P(s') ds'$. The integration limit for the
first 2 quantities is chosen as $s_0= 1.4$ which approximates the
common crossing point\cite{HofS94b} of all LSD curves as can be
seen in Fig.\ \ref{fig-lsd-LSDshift}.
Thus $P(s_0)$ is independent of the distance $|z-z_{\rm c}|$ to
the critical point and the system size $N$. Note that $N$ is
directly related to the RG step $n$ by $N=2^n$.
The double integration in $\alpha_{\rm I}$ is numerically
advantageous since fluctuations in $P(s)$ are smoothed.
One can now apply the finite-size-scaling approach from Eq.\
(\ref{eq-lsd-scal}) for $\alpha_{\rm P,I,S}(N,z_0)$. Since
$\alpha_{\rm P,I,S}(N,z_0)$ is analytical for finite $N$, one can
expand the scaling function $f$ at the critical point. The
first-order approximation yields
\begin{equation}
  \alpha(N,z_0)\sim \alpha(N,z_{\rm c})+a |z_0-z_{\rm c}| N^{1/\nu}
\end{equation}
where $a$ is a coefficient. Better results are obtained using a
higher-order expansion proposed by Slevin and
Ohtsuki\cite{SleO99a}, but contributions from an irrelevant
scaling variable can be neglected since the transition point
$z_0=0$ is known \cite{CaiRR03}.

\begin{figure}[!tbh]
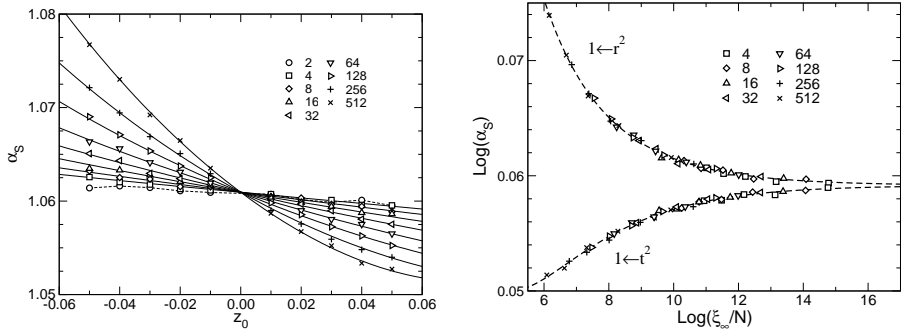

\vspace{0.5cm}
  \centerline{\includegraphics[width=0.45\columnwidth]{Alpha_S.func.eps}
\quad
\includegraphics[width=0.45\columnwidth]{Alpha_S.FSS.eps}}
\caption{\label{fig-lsd-AlphaS} Behavior of $\alpha_{\rm S}$ at
the QH transition. Left: Data for RG iterations $n=1,\ldots,9$
corresponding to
  effective system sizes $N=2^n=2,\ldots,512$. Full lines indicate the
  functional dependence according to FSS using the $\chi^2$
 minimization with ${\cal O}_{\nu}=2$ and ${\cal O}_{z}=3$. Because
  of large deviations data for $N=2$ were excluded from the fitting.
  Right: Resulting FSS curves.  Different symbols correspond to
  different effective system sizes $N=2^n$.}
\end{figure}

In order to obtain the functional form of $f$ the fitting
parameters, including $\nu$, are evaluated by a nonlinear
least-square ($\chi^2$) optimization. In Fig.\
\ref{fig-lsd-AlphaS} the resulting fits for, e.g., $\alpha_{\rm
S}$ at the transition are shown. The fits are chosen such that the
total number of parameters is kept small and the fit agrees well
with the numerical data. The corresponding scaling curve is
displayed in the right panel of Fig.\ \ref{fig-lsd-AlphaS}. In the
plots the two branches for complete reflection ($z_0<0$) and
complete transmission ($z_0>0$) can be distinguished clearly. In
order to estimate the error of the fitting procedure the over
$100$ results for $\nu$ obtained by different orders of the
expansion, system sizes $N$, and $z$ ranges around the transition
are compared \cite{CaiRR03,Cai04}.
The value of $\nu=2.37\pm 0.02$ is calculated as average over many
individual fits for $\alpha_{\rm P,I}$. It is in excellent
agreement with $2.39\pm0.01$ calculated in section
\ref{sec-pglsd-critexp}. The value obtained from the
parameter-free scaling quantity $\alpha_{\rm S}=\int_0^\infty s^2
P(s) ds$\cite{ZhaK95b} is $\nu=2.33\pm0.05$, but scaling is less
reliable due to the insufficient statistics in the large-$s$ tail.

Finally the influence of the initial conditions on the result of
the LSD and the one-parameter scaling has been studied in Ref.\
\cite{CaiRR03}. The corresponding LSD agrees with GUE less
convincingly than the LSD computed using the true $P_{\rm c}(G)$.
Estimates of the critical exponent yield also less accurate values
$\nu_{\rm I}=2.43\pm0.02$ and $\nu_{\rm P}=2.46\pm0.03$ which are
nevertheless still reasonably close to $\nu=2.37\pm0.02$. Thus our
choice of the initial distribution $P_{\rm c}(G)$ is indeed valid.

\section{Extensions of the RG approach}
\label{sec-rhmacro}

\subsection{The QH-to-insulator transition}
\label{sec-rhmacro-rh}

\subsubsection{Evolution of the resistance distributions $P(R_{\rm L})$ and $P(R_{\rm H})$}
\label{sec-rhmacro-rh-dist}

\begin{figure}[tbh]
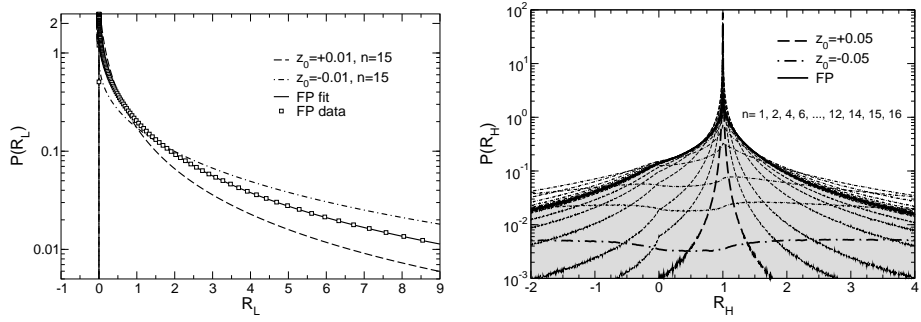

  \centerline{\includegraphics[width=0.455\columnwidth]{fig-P_RL_linlog.eps}
\quad
\includegraphics[width=0.46\columnwidth]{fig-P_RH-new.eps}}
\caption{\label{fig-rhall-5SP-z0} Left: Distribution of the
longitudinal resistance $R_{\rm L}$. The $\Box$ symbols indicate
the FP distribution, the solid line is a fit to a log-normal
distribution and the dot-dashed and dashed lines show the
distribution after $n=15$ RG iterations into the conductance
regimes $G(z_0 < 0)\rightarrow 0$ and $G(z_0 > 0) \rightarrow 1$.
Only every 5th data point is shown for $P(R_{\rm L})$.
Right: Distribution of the Hall resistance $R_{\rm H}$.  The FP
  distribution has been shaded to zero. For the $G(z_0 < 0)\rightarrow
  0$ and $G(z_0 > 0) \rightarrow 1$ regimes, we have indicated the
  distributions after $n=16$ RG iterations by bold dashed and
  dot-dashed lines.  }
\end{figure}

In order to model the transition into the insulating regime, we
shift the initial distribution $Q_0(z) \rightarrow Q_0(z-z_0)$ by
a small $z_0$. For $z_0 < 0$ and $z_0 > 0$, the RG flow will then
drive the distributions into the insulating, $G \rightarrow 0$,
and plateau, $G\rightarrow 1$, regimes, respectively. In Fig.\
\ref{fig-rhall-5SP-z0}, we show the resulting distributions after
many RG steps.
For $P(R_{\rm L})$, we see in Fig.\ \ref{fig-rhall-5SP-z0} that
the flow towards $G\rightarrow 1$ results in a decrease of large
$R_{\rm L}$ events, whereas conversely, the regime $G \rightarrow
0$ leads to an increase in large $R_{\rm L}$ values and a decrease
of the maximum value in $P(R_{\rm L})$.
For $P(R_{\rm H})$, Fig.\ \ref{fig-rhall-5SP-z0} shows that the
plateau regime $G\rightarrow 1$ gives a highly singular peak at
the dimensionless quantized Hall value $1$, corresponding to a
perfect Hall plateau. On the other hand, the insulating regime $G
\rightarrow 0$ shows an increase of weight in the tails of
$P(R_{\rm H})$ and the eventual obliteration of any central peak.
%
%
Very large absolute values of $R_{\rm L}$ and $R_{\rm H}$ in the
tails are obviously the consequence of a very small denominator
$t'^2$ in Eqs.\ (\ref{eq-RL}) and (\ref{eq-rhall-rg}) meaning that
the current is almost totally reflected by the RG unit. This
scenario is possible since $P_n(t^2)$ also exhibits strong
fluctuations and spreads over the whole range between $0$ and $1$.

\subsubsection{Evolution of the average resistances}
\label{sec-rhmacro-rh-average}

In order to extract an averaged $R_{\rm L}$ and $R_{\rm H}$ from
these non-standard distributions, we should now select an
appropriate mean $\langle \cdot \rangle$ that characterizes and
captures the essential physics and allows comparison with the
experimental data. The precise operational definition is also
important as it corresponds to different possible experimental
setups. Therefore, we consider several means: (i) arithmetic
$\langle R \rangle_{\rm ari}= \sum_{i} R_i/N$, (ii)
geometric/typical $\langle R \rangle_{\rm typ}= \exp \sum_{i} \ln
R_i/N$, (iii) median (central value) $\langle R \rangle_{\rm
med}$, where $N$ denotes the number of samples ($\gtrsim 10^8$) in
each case. The median and the typical mean (and their variances
\cite{ZulS01}) are less sensitive to extreme values than other
means (such as, e.g.\ root-mean-square and harmonic mean) and this
makes them a better measure for highly skewed and long-tailed
distributions such as $P(R_{\rm L})$ and $P(R_{\rm H})$ in the
insulating regime. In the plateau regimes, the distributions are
less skewed, particularly for $R_{\rm H}$ and the difference in
the means becomes less important.

We are left with determining in which operational order to apply
the averaging procedure. For $R_{\rm L}$, as measured via Ohm's
law (\ref{eq-RL}) as a ratio, it is obvious that the appropriate
average should be ${R}_{\rm L} = \frac{1}{\langle G \rangle} - 1$
(and not $\langle\frac{1}{G}\rangle -1$). For $R_{\rm H}$, the
situation is less straightforward due to the definition of $U_{\rm
H}$ in (\ref{eq-rhall-rg}). A simple average is $\langle U_{\rm H}
\rangle$, i.e. using the appropriate $P(R_{\rm H})$. Similar to
the experimental procedure, we can also estimate $U_{\rm H}$ via
$\langle a_{u} - a_{v} \rangle$ for each $B$ field direction
separately. In Ref.\ \cite{ZulS01}, it has been suggested that a
more appropriate average $\langle U_{\rm H}\rangle^{*}$ can be
constructed from $\langle a_{u} \rangle - \langle a_{v}
\rangle$. This later procedure corresponds to measuring the
voltage drop between positions $a_{u}$ and $a_{v}$ in Fig.\
\ref{fig-rgstruct} by separately measuring the individual voltages
with respect to ground and then recombining them. For $R_{\rm H}$,
this yields
\begin{equation}
\label{eq-rhall-rhtyp}
  \langle R_{\rm H}\rangle_{\rm typ}^*=\frac{1}{2} \frac{\left(\langle a_u^{(B)}\rangle_{\rm
  typ}-\langle a_v^{(B)}\rangle_{\rm typ}\right) - \left(\langle
  a_u^{(-B)}\rangle_{\rm typ}-\langle a_v^{(-B)}\rangle_{\rm
  typ}\right)} {\langle t'^2\rangle_{\rm typ}} .
\end{equation}
\begin{figure}[tbh]
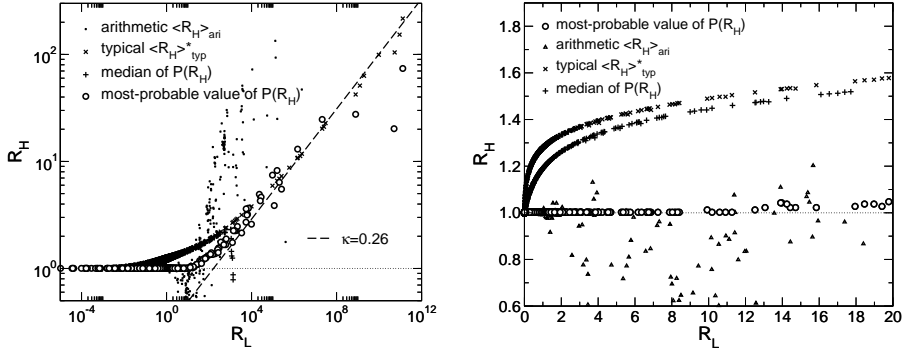

\centerline{\includegraphics[width=0.45\columnwidth]{fig-RHRL-loglog-new_mod.eps}
\quad
\includegraphics[width=0.45\columnwidth]{fig-RHRL-linlin-new_mod.eps}}
\caption{\label{fig-RHRL-loglog}
 Left: Dependence of averaged $R_{\rm H}$ on averaged $R_{\rm L}$ for various means obtained from $10^8$ samples.
The dashed line describes the divergence $R_{\rm H} \propto R_{\rm
L}^{\kappa}$ of the typical mean $\langle R_{\rm
H}\rangle^{*}_{\rm typ}$, the median $\langle R_{\rm
H}\rangle_{\rm med}$ and the most probable value of $P(R_{\rm
H})$.
  The variance of the arithmetic mean diverges as $R_{\rm L}\rightarrow \infty$ and
  $\langle R_{\rm H}\rangle_{\rm ari}$ is no longer a useful characteristic of
  $P(R_{\rm H})$.
  The horizontal dotted line indicates $R_{\rm H}=1$.
Right:\label{fig-RHRL-linlin}
  Plot of the same data as in Fig.\ \protect\ref{fig-RHRL-loglog} but on
  a linear scale and for experimentally accessible resistance values.
  Here the almost quantized behavior for the most probable value of $P(R_{\rm H})$
  becomes even more pronounced.
The horizontal dotted line indicates $R_{\rm H}=1$.}
\end{figure}
In Fig.\ \ref{fig-RHRL-loglog} we show the resulting dependence
$R_{\rm H}(R_{\rm L})$ when these averaging definitions are being
used.
In the plateau regime $R_{\rm L}<1$, the nearly constant behavior
of all $R_{\rm H}$ averages is as expected. For $R_{\rm L}
\rightarrow \infty$, we get a divergent $R_{\rm H}$ when using the
median and typical means as suggested by Refs.\
\cite{PryA99,ZulS01} to compute both $R_{\rm L}$ and $R_{\rm H}$.
This divergence can be captured by a power-law $R_{\rm H} \propto
R_{\rm L}^{\kappa}$ with $\kappa \approx 0.26$.
The arithmetic mean for large $R_{\rm L} \gg 1$ quickly becomes
instable and no useful information can be inferred.

Reducing the information to the experimentally more relevant
resistance regime of a few times $h/e^2$, we replot the $R_{\rm
L}$ and $R_{\rm H}$ data in Fig.\ \ref{fig-RHRL-linlin} (right
panel) on a linear scale. In addition to the three means above, we
also show the behavior of the most-probable value $\hat{R}_{\rm
H}$ at which $P(R_{\rm H})$ has a maximum. This estimate
$\hat{R}_{\rm H}(R_{\rm L})$ appears relevant in the experimental
setup where $10^{8}$ different samples cannot be easily measured
and the full distribution functions cannot be constructed in
similar detail.
Most importantly, for the range of $R_{\rm L}$ values shown in
Fig.\ \ref{fig-RHRL-linlin}, the value of $\hat{R}_{\rm H}$
deviates only slightly from its quantized value $1$ at the
transition and $R_{\rm L}+R_{\rm H}\approx R_{\rm 2t}$
\cite{PelSCS03}. Therefore, the experimental estimate of $R_{\rm
H}$ appears to support the notion of the quantized Hall insulator.
Indeed, the deviations from $1$ are less than $10\%$ until $R_{\rm
L} \sim 25$. However, going back to the left panel of Fig.\
\ref{fig-RHRL-loglog}, we see that in the strongly insulating
regime $R_{\rm L} \rightarrow \infty$ also $\hat{R}_{\rm H}$
diverges with a power-law that is well-described by $\langle R
\rangle^*_{\rm typ}$. We emphasize that fluctuations in
$\hat{R}_{\rm H}$ for large ${R}_{\rm L} \gg 10^5$ are due to
numerical inaccuracies in $P(R_{\rm H})$ and decrease upon further
increasing the number of samples.

\begin{figure}[tbh]
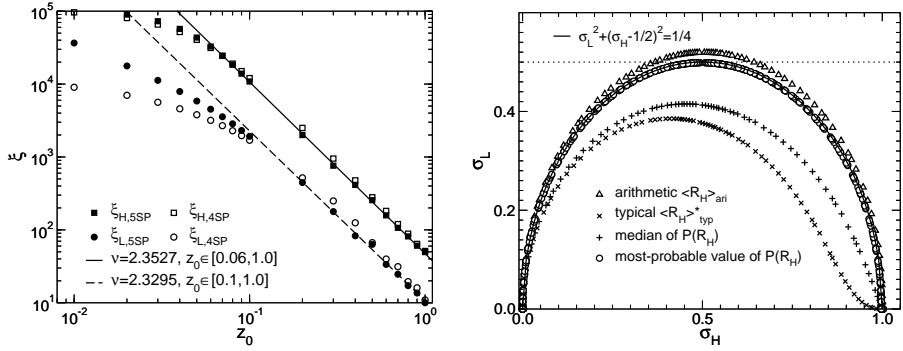

 \centerline{\includegraphics[width=0.45\columnwidth]{fig-xi_z0_loglog.eps}
\quad\includegraphics[width=0.45\columnwidth]{fig-SxxSxy_linlin_new_mod.eps}}
\caption{\label{fig-xi-z0-loglog}
  Left: Dependence of the localization lengths $\xi_{L}$ and $\xi_{H}$ on
  the initial shift $z_0$ using the ansatz (8) of \cite{PryA99}. We
  obtain a reasonable estimate of the critical exponent $\nu$.
Right: \label{fig-SxxSxy-linlin}
  The semi circle law for different means as in Fig.\ \ref{fig-RHRL-linlin}.
  The most probable values ($\circ$) show perfect semi-circle (solid line) behavior.
  The thin dotted line denotes \protect{$\sigma_{\rm L}={1}/{2}$}.}
\end{figure}

The results for $R_{\rm L}$ and $R_{\rm H}$ in the localized
regime can be very well described by an exponential scaling
function with finite-size correction \cite{PryA99}
$R_{\rm L,H}(2^n,z) \propto 2^{\gamma n} \exp [ {2^n}{\xi^{-1}_{\rm L,H}(z)}]$.
Plotting $\xi_{\rm L,H}$ as a function of small perturbation
$z_0$, we find $\xi_{\rm L,H}(z_0) \propto z_0^{-\nu_{\rm L,H}}$
with $\nu_{\rm L,H} \approx 2.35$ as shown in Fig.\
\ref{fig-xi-z0-loglog}. Thus we recover the universal divergence
of the localization length $\xi$ even when using the RG in the
insulating regime \cite{Huc92}.
An equally reliable estimate of the irrelevant exponent $\gamma$
appears not possible for our data.
As usual, the 4-terminal resistances can be converted into the
respective conductances via $\sigma_{\rm L,H} = {R_{\rm
    L,H}}/\left({R_{\rm L}^2 + R_{\rm H}^2}\right)$.  For these
conductances, one expects the semi-circle law $\sigma_{\rm L}^2 +
\left(\sigma_{\rm H}-{1}/{2}\right)^2 = \left( {1}/{2} \right)^2$ to
hold \cite{RuzF95}. In Fig.\ \ref{fig-SxxSxy-linlin}, we show that the
most-probable values capture the overall shape and symmetry properties
best \cite{HilSST98,PelSCS03}, the other averages show pronounced
deviations. We also note that the relation $R_{\rm L}(z_0)=1/R_{\rm
  L}(-z_0)$ is obeyed by all means \cite{ShaHLT98}. This is a
consequence of the reflection symmetry of $P(G)$ \cite{CaiRSR01}.

\subsection{Macroscopic inhomogeneities in the RG approach}
\label{sec-rhmacro-macro}

\subsubsection{Implementing power-law correlations in the RG approach}
\label{sec-rhmacro-macro-implement}

A natural way to incorporate a quenched disorder into the CC model
is to ascribe a certain random shift, $z_Q$, to each SP height,
and to assume that the shifts at different SP positions, ${\bf r}$
and ${\bf r}'$, are correlated as
\begin{equation}
\label{eq-macros-corr} \langle z_Q({\bf r})z_Q({\bf r}')\rangle
\propto |{\bf r}-{\bf r}'|^{-\alpha},
\end{equation}
with $\alpha >0$. After this, the conventional transfer-matrix
methods of \cite{FisL81,MacK81} could be employed for numerically
precise determination of $\langle G \rangle$, the distribution
$P_{\rm c}(G)$, its moments, and most importantly, the critical
exponent, $\nu$.  However, the transfer-matrix approach for a 2D
sample is usually limited to fairly small cross sections (e.g., up
to $128$ in \cite{JovW98}) due to the numerical complexity of the
calculation. Therefore, the spatial decay of the power-law
correlation by, say, more than an order in magnitude is hard to
investigate for small $\alpha$.

On the other hand, the RG approach is perfectly suited to study
the role of the quenched disorder.  First, after each step of the
RG procedure, the effective system size doubles.  At the same
time, the magnitude of the smooth potential, corresponding to the
spatial scale $r$, falls off with $r$ as $r^{-\alpha/2}$. As a
result, the modification of the RG procedure due to the presence
of the quenched disorder reduces to a scaling of the disorder
magnitude by a {\em constant} factor $2^{-\alpha/2}$ at each RG
step.  Second, the RG approach operates with the conductance
distribution $P_n(G)$ which carries information about {\em all}
the realizations of the quenched disorder within a sample of size
$2^n$.  This is in contrast to the transfer-matrix
approach\cite{FisL81,MacK81}, within which a small increase of the
system size requires the averaging over the quenched disorder
realizations to be conducted again.

The above consideration suggests the following algorithm. For the
homogeneous case all SPs constituting the new super-SP are assumed
to be identical, which means that the distribution of heights,
$Q_n(z)$, for all of them is the same.  For the correlated case
these SPs are no longer identical, but rather their heights are
randomly shifted by the long-ranged potential.  In order to
incorporate this potential into the RG scheme, $Q_n(z_i)$ should
be replaced by $Q_n(z_i-\Delta^{(n)}_i)$ for each SP, $i$, where
$\Delta^{(n)}_i$ is the random shift. Then the power-law
correlation of the quenched disorder enters into the RG procedure
through the distribution of $\Delta^{(n)}_i$. That is, for each
$n$ the distribution is Gaussian with the width $\beta
(2^n)^{-\alpha/2}$.  For large enough $n$ the critical behavior
should not depend on the magnitude of the correlation strength
$\beta$, but on the power, $\alpha$, only.
%
\subsubsection{A new critical behavior} \label{sec-macros-result}

\begin{figure}[tbh]
\centerline{\includegraphics[width=0.45\columnwidth]{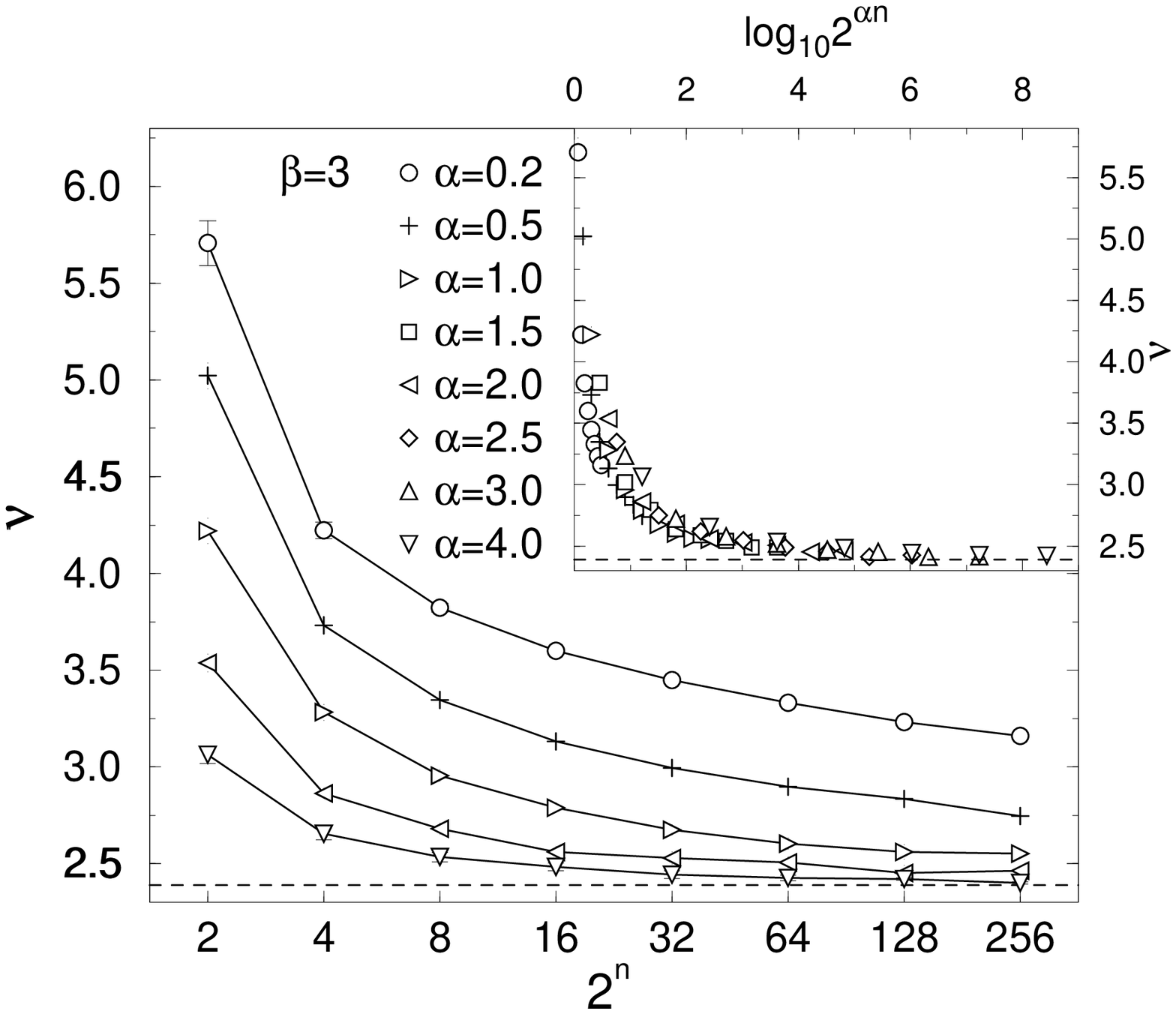}
\quad \includegraphics[width=0.45\columnwidth]{nu_alpha.eps}
} \caption{\label{fig-macros-nu-b3} Left: Critical exponent $\nu$
obtained by the QH-RG approach as a function
  of RG scale $2^n$ for $\beta=3$ and different correlation exponents
  $\alpha$.  The dashed line indicates $\nu=2.39$, which is the value
  obtained for uncorrelated disorder. For clarity, the errors
  are shown only for $\alpha=0.2$ and $4$.  Inset: $\nu$ vs $2^{\alpha
    n/2}$ shows deviations from scaling.
  \label{fig-macros-alpha} Right: Dependence of the critical
  exponent $\nu$ on correlation exponent $\alpha$ for different
  $\beta=1, 2, 3$, and $4$ as obtained after eight QH-RG iterations.
  The dotted line indicates $\alpha_{\rm c}=0.84$. The dot-dashed line
  $\nu=2/\alpha$ for $\alpha<\alpha_{\rm c}$ follows from the
  extended Harris criterion \protect\cite{WeiH83} for
  classical percolation.}
\end{figure}

We find that for all values of $\alpha>0$ in correlator
(\ref{eq-macros-corr}) the FP distribution is identical to the
uncorrelated case within the accuracy of the calculation. In
particular, $\langle G \rangle=0.498$ is unchanged. However, the
convergence to the FP is numerically less stable than for
uncorrelated disorder due to the correlation-induced broadening of
$Q_n(z)$ during each iteration step.
In order to compute the critical exponent $\nu(\alpha)$ the RG
procedure is started from $Q_0(z-z_0)$, as in the uncorrelated
case, but in addition the random shifts are incorporated. As
already explained these shifts are a result of the quenched
disorder in generating the distribution of $z$ at each RG step.
The outcome is shown in Fig.\ \ref{fig-macros-nu-b3}. It
illustrates that for increasing long-ranged character of the
correlation (decreasing $\alpha$) the convergence to a limiting
value of $\nu$ slows down drastically.  Even after eight RG steps
(i.e., a magnification of the system size by a factor of $256$),
the value of $\nu$ with long-ranged correlations still differs
appreciably from $\nu=2.39$ obtained for the uncorrelated case.
The RG procedure becomes unstable after eight iterations, i.e.,
$z_{\rm{max},9}$ can no longer be obtained reliably from $Q_9(z)$.
Unfortunately, for small $\alpha$ the convergence is too slow to
yield the limiting value of $\nu$ after eight steps only.  For
this reason, the above method is, strictly speaking, unable to
unambiguously answer the question whether sufficiently long-ranged
correlations result in an $\alpha$-dependent critical exponent
$\nu(\alpha)$, or whether the value of $\nu$ eventually approaches
the uncorrelated value of $2.39$.  Nevertheless, the results in
Fig.\ \ref{fig-macros-nu-b3} indicate that the effective critical
exponent exhibits a sensitivity to the long-ranged correlations
even after a large magnification by $256\times 256$. Since this
factor coincides with the change in scale from the microscopic
magnetic length to the realistic samples with finite sizes of
several $\mu{\rm m}$, macroscopic inhomogeneities are able to
affect the results of scaling studies.
%
%
Note further that as shown in the inset of Fig.\
\ref{fig-macros-nu-b3} one can observe scaling of $\nu$ values
when plotted as function of a renormalized system size $2^{\alpha
n/2}$ only for large correlation strength $\beta$. In this case
the disorder broadening during the first RG steps becomes
comparable to the FP distribution. Therefore in the QH-RG the
correlated disorder dominates over the initial FP distribution.
One should emphasize that $\nu(\alpha)$ obtained after eight RG
steps always {\em exceeds} the uncorrelated value.

Figure \ref{fig-macros-alpha} shows the values of $\nu$ obtained
after the eighth RG step as a function of the correlation exponent
$\alpha$ for different dimensionless strengths $\beta$ of the
quenched disorder.  It is seen that in the domain of $\alpha$,
where the values of $\nu$ differ noticeably from $\nu=2.39$, their
dependence on $\beta$ is strong.  According to the extended Harris
criterion \cite{WeiH83,Wei84}, $\nu(\alpha)$ is expected to
exhibit a cusp at the $\beta$-independent value $\alpha_c=2/2.39
\approx 0.84$. From the results in Fig.\ \ref{fig-macros-alpha},
two basic observations can be made. For a small enough magnitude
of the long-range disorder, one observes a smooth enhancement of
$\nu(\alpha)$ with decreasing $\alpha$ without a cusp.  Although
such a behavior might be caused by the relatively small number of
RG steps, the data can be relevant for realistic samples which
have a finite size and a finite phase-breaking length governed by
temperature.
At the largest $\beta=4$, there is still no clear cusp but the
$\alpha$-dependence for $\alpha<\alpha_{\rm c}$ is closer to the
$\nu=2/\alpha$ predicted by the extended Harris criterion.
Unfortunately, the numerics becomes progressively unstable,
forbidding to go to even larger $\beta$.

\subsubsection{Intrinsic short-range disorder in quantum
  percolation} \label{sec-macros-short}

Here one should note that there is a crucial distinction between the
classical case and the quantum regime of the electron motion considered
in the present RG approach.
In this study, the correlation of the {\em heights} of the SPs has
been incorporated into the RG scheme. At the same time it was
assumed that the {\em Aharonov-Bohm phases} acquired by an
electron upon traversing the neighboring loops are {\em
uncorrelated}.  This assumption implies that, in addition to the
long-ranged potential, a certain short-ranged disorder causing a
spread in the perimeters of neighboring loops of the order of the
magnetic length is present in the sample. The consequence of this
short-range disorder is the sensitivity of the results to the
value of $\beta$ which parametrizes the magnitude of the
correlated potential.  The presence of this short-range disorder
affecting exclusively the Aharonov-Bohm phases significantly
complicates the observation of the cusp in the $\nu (\alpha)$
dependence at $\alpha \approx 0.84$, as is expected from the
extended Harris criterion. A more detailed investigation of this
effect has been given in Ref.\ \cite{CaiRSR01}. We emphasize that
an unambiguous demonstration of the validity of the extended
Harris criterion has recently been given by the Liouville approach
to the random Landau matrix model \cite{SanMK03}.

\section{Conclusions}
 \label{sec-concl}

Real-space RG approaches to problems of quantum statistical
physics have been applied with much success in many context. The
RG necessarily uses only a local approximation to the full
connectivity of the network.  Thus it is important to choose the
RG structures such that the overall symmetries of the network are
retained. In the present work, we have shown various examples
where such a strategy is permissible in the context of the integer
QH effect. Particularly for the determination of critical
distributions $P_{\rm c}(G)$, $P_{\rm c}(R_{\rm L})$, $P_{\rm
c}(R_{\rm H})$ and the critical exponent $\nu=2.39\pm 0.01$, this
allowed us to study of the critical properties of the QH
transition with surprisingly high accuracy.
It seems possible to further enhance the accuracy of the RG by
increasing the size of the RG unit \cite{WeyJ98}. However, in
order to fully benefit from the increased size, it is essential
that an analytic solution as in Eq.\ (\ref{eq-rg-qhrg}) is
obtained. Otherwise, numerical round-off and stability effects
will lead to an overall decreased accuracy. We emphasize also that
in the case of larger RG units the size-rescaling factor needs to
be calculated with care.

The robustness of the 5SP RG when computing universal properties
is moreover demonstrated by an FSS analysis of the LSD around the
QH transition. The exponent $\nu=2.37\pm 0.02 $ of the
localization length obtained by a nonlinear $\chi^2$ minimization
is in excellent agreement with the value reported in the
literature. It is surprisingly good when keeping in mind that it
was derived just from the four loops of the RG unit which
therefore seem to capture the essential physics of the QH
transition.  The success of the  RG approach to the LSD can be
attributed to
the description of the transmission amplitudes $t$ of the SPs by a
full distribution function $P(t)$ while for network models usually
a fixed value $t(E)$ is assigned to all SPs for simplicity.
Also, the phases are associated with full loops in the network and
not with single SP-SP links. Due to these reasons the RG iteration
is always governed by the quantum critical point of the QH
transition.  Even when starting the iteration with a distribution
$P_0(G)$ totally different from $P_{\rm c}(G)$, but still
symmetric with respect to $G=0.5$, one approaches the same results
after a few iterations.

Having established that the RG approach works at the QH
transition, we then set out to extend the method to distributions
away from criticality. We have shown in a quantum coherent
calculation that the insulating Hall behavior of $R_{\rm H}(R_{\rm
L}\rightarrow\infty)$ is dominated by the power-law divergence
$R_{\rm H} \propto R_{\rm L}^{\kappa}$. However, in the
experimentally attained range $R_{\rm L}\leq 10$
\cite{HilSST98,LanPVP02,PelSCS03}, the deviation from a quantized
$R_{\rm H}=1$ is very small and the onset of the divergence for
$R_{\rm L}\gg 10$ is yet to be explored.

It was argued for a long time that the enhanced value of the
critical exponent $\nu$ extracted from the narrowing of the
transition region with temperature \cite{WeiLTP92,SchVOW00} has
its origin in the long-ranged disorder present in GaAs-based
samples. To our knowledge, the present RG approach was the first
attempt to quantify this argument. As a result the random
potential with a power-law correlator leads to values of $\nu$
exceeding $\nu \approx 2.35$.

The result indicates that macroscopic inhomogeneities must lead to
smaller values of $\kappa\propto 1/\nu$.  Experimentally, the
value of $\kappa$ smaller than $0.42$ was reported in a number of
early (see, e.g., \cite{KocHKP92} and references therein) as well
as recent \cite{Lan00} works. This fact was accounted for by
different reasons (such as temperature dependence of the phase
breaking time, incomplete spin resolution, valley degeneracy in
Si-based MOSFETs, and inhomogeneity of the carrier concentration
in GaAs-based structures with a wide spacer).  Briefly, the spread
of the $\kappa$ values was attributed to the fact that the
temperatures were not low enough to assess the truly critical
regime.  The possibility of having $\kappa < 0.4$ due to the
correlation-induced dependence of the effective $\nu$ on the
phase-breaking length or, ultimately, on the sample size, as in
Fig.\ \ref{fig-macros-nu-b3}, was not considered.

It should be pointed out that the limited number of RG steps permitted
by the numerics nevertheless allows one to trace the evolution of the
wave functions from {\em microscopic} scales (of the order of the
magnetic length) to {\em macroscopic} scales (of the order of several $
\mu{\rm m}$) which are comparable to the sizes of the samples used in
the experimental studies of scaling (see e.g.,
\cite{KocHKP91b,KocHKP92}) and much larger than the
samples\cite{CobBF99,CobK96} used for the studies of mesoscopic
fluctuations.
Another {\em qualitative} conclusion of this study is that the
spatial scale at which the exponent $\nu$ assumes its
``infinite-sample'' value is much larger in the presence of the
quenched disorder than in the uncorrelated case. In fact this
scale can be of the order of microns.  This conclusion can also
have serious experimental implications. That is, even if the
sample size is much larger than this characteristic scale, this
scale can still exceed the phase-breaking length, which would mask
the true critical behavior at the QH transition.

Let us briefly comment on the possible future application of the
method. An obvious possibility is the application to extended
variants of the standard CC model, e.g. a network with at least
two channels per link in order to describe the mixing of Landau
levels \cite{WanLW94}. There remain several interesting questions
such as the behavior of the critical properties at the QH
transition when changing from strong to weak magnetic field. This
case could be modelled by bi-directional links in the network,
which would allow one to trace the transition from the
universality class GUE to GOE.
And last, but certainly not least, we have described the integer
QH effect within a non-interacting electron picture, but
experimental results clearly indicate the influence of
interactions \cite{EngSKT93,PanSTW97}. Because a full treatment of
many-body interactions is rather difficult one might consider in
an approximate view only a few interacting particles. In this
approach the two-interacting particle problem is reduced to a
single-particle problem by increasing the effective spatial
dimension and including long-range correlations in the disorder
potential\cite{She94}. Concerning the CC model one has to
construct an effective four-dimensional network. From this new CC
network a suitable RG unit should be extracted. This tasks are by
no means trivial but first steps have already been undertaken
\cite{ApaR03}.

\section*{Acknowledgements}

It is a pleasure to thank R.\ Ball, B.\ Huckestein, R.\ Klesse,
M.\ Raikh, M.\ Schreiber, and U.\ Z\"{u}licke for stimulating
discussions. Financial support by the DFG priority research
program "Quanten-Hall-Systeme", EPSRC and DFG SFB393 is gratefully
acknowledged.



\begin{thebibliography}{100}

\bibitem{KliDP80}
K.~v. Klitzing, G. Dorda, and M. Pepper, Phys. Rev. Lett. {\bf 45},  494
  (1980).

\bibitem{TsuSG82}
D.~C. Tsui, H.~L. {St\"{o}rmer}, and A.~C. Gossard, Phys. Rev. Lett. {\bf 48},
  1559  (1982).

\bibitem{ChaP95}
T. Chakraborty and P. {Pietil\"{a}nen}, {\em The Quantum {Hall} Effects}
  (Springer, Berlin, 1995).

\bibitem{Yos02}
D. Yoshioka, {\em The Quantum {Hall} Effect}, {\em Springer Series in
  Solid-State Sciences 133} (Springer, Berlin, 2002).

\bibitem{AokA81}
H. Aoki and T. Ando, Solid State Commun. {\bf 38},  1079  (1981).

\bibitem{JanVFH94}
M. Janssen, O. Viehweger, U. Fastenrath, and J. Hajdu, {\em Introduction to the
  {Theory} of the {Integer} {Quantum} {Hall} effect} (VCH, Weinheim, 1994).

\bibitem{Lau81}
R.~B. Laughlin, Phys. Rev. B {\bf 23},  5632  (1981).

\bibitem{Pra81}
R.~E. Prange, Phys. Rev. B {\bf 23},  4802  (1981).

\bibitem{Pru84}
A.~M.~M. Pruisken, Nucl. Phys. B {\bf 235},  277  (1984).

\bibitem{ThoKNN82}
D.~J. Thouless, M. Kohmoto, M.~P. Nightingale, and M.~d. Nijs, Phys. Rev. Lett.
  {\bf 49},  405  (1982).

\bibitem{Huc95}
B. Huckestein, Rev. Mod. Phys. {\bf 67},  357  (1995).

\bibitem{ChaC88}
J.~T. Chalker and P.~D. Coddington, J. Phys.: Condens. Matter {\bf 21},  2665
  (1988).

\bibitem{BatS96}
M. Batsch and L. Schweitzer,  in {\em High Magnetic Fields in Physics of
  Semiconductors II: Proceedings of the International Conference, {W\"urzburg}
  1996}, edited by G. Landwehr and W. Ossau (World Scientific Publishers Co.,
  Singapore, 1997), pp.\ 47--50, {ArXiv}: cond-mat/9608148.

\bibitem{Huc92}
B. Huckestein, Europhys. Lett. {\bf 20},  451  (1992).

\bibitem{KleM97}
R. Klesse and M. Metzler, Phys. Rev. Lett. {\bf 79},  721  (1997).

\bibitem{LeeWK93}
D.-H. Lee, Z. Wang, and S. Kivelson, Phys. Rev. Lett. {\bf 70},  4130  (1993).

\bibitem{Met98b}
M. Metzler, J. Phys. Soc. Japan {\bf 67},  4006  (1998).

\bibitem{AroJS97}
D.~P. Arovas, M. Janssen, and B. Shapiro, Phys. Rev. B {\bf 56},  4751  (1997),
  {ArXiv}: cond-mat/9702146.

\bibitem{GalR97}
A.~G. Galstyan and M.~E. Raikh, Phys. Rev. B {\bf 56},  1422  (1997).

\bibitem{BinDFN92}
J.~J. Binney, N.~J. Dowrick, A.~J. Fisher, and M.~E.~J. Newman, {\em The Theory
  of Critical Phenomena: An Introduction to the Renormalization Group} (Oxford
  University Press, Oxford, UK, 1992).

\bibitem{Wil83}
K.~G. Wilson, Rev. Mod. Phys. {\bf 55},  583  (1983).

\bibitem{Car96}
J.~L. Cardy, {\em Scaling and Renormalization in Statistical Physics}
  (Cambridge University Press, Cambridge, 1996).

\bibitem{Sal99}
M. Salmhofer, {\em Renormalization: an Introduction} (Springer, Berlin, 1999).

\bibitem{GolWSS93}
V.~J. Goldman, J.~K. Wang, B. Su, and M. Shayegan, Phys. Rev. Lett. {\bf 70},
  647  (1993).

\bibitem{HilSST98}
M. Hilke {\it et~al.}, Nature {\bf 395},  675  (1998).

\bibitem{LanPVP02}
D.~d. Lang {\it et~al.}, Physica E {\bf 12},  666  (2002).

\bibitem{ShaTSC97}
D. Shahar {\it et~al.}, Solid State Commun. {\bf 102},  817  (1997).

\bibitem{ShaTSS96}
D. Shahar {\it et~al.}, Science {\bf 274},  589  (1996).

\bibitem{HugNFL94}
R. Hughes {\it et~al.}, J. Phys.: Condens. Matter {\bf 6},  4763  (1994).

\bibitem{MurHLC00}
S.~Q. Murphy {\it et~al.}, Physica E {\bf 6},  293  (2000).

\bibitem{PanSTW97}
W. Pan {\it et~al.}, Phys. Rev. B {\bf 55},  15431  (1997).

\bibitem{ShaTSB95}
D. Shahar {\it et~al.}, Phys. Rev. Lett. {\bf 74},  4511  (1995).

\bibitem{ShaTSS97}
D. Shahar {\it et~al.}, Phys. Rev. Lett. {\bf 79},  479  (1997).

\bibitem{SchVOW00}
R.~T. F.~v. Schaijk {\it et~al.}, Phys. Rev. Lett. {\bf 84},  1567  (2000).

\bibitem{Shi04}
E. Shimshoni, Mod. Phys. Lett. B {\bf 18},  923  (2004).

\bibitem{PelSCS03}
E. Peled {\it et~al.}, Phys. Rev. Lett. {\bf 90},  246802  (2003).

\bibitem{PryA99}
L.~P. Pryadko and A. Auerbach, Phys. Rev. Lett. {\bf 82},  1253  (1999).

\bibitem{ZulS01}
U. {Z\"{u}licke} and E. Shimshoni, Phys. Rev. B {\bf 63},  241301  (2001),
  {ArXiv}: cond-mat/0101443.

\bibitem{EveMPW99}
F. Evers, A.~D. Mirlin, D.~G. Polyakov, and P. {W{\"o}lfle}, Phys. Rev. B {\bf
  60},  8951  (1999).

\bibitem{CooHHR97}
N.~R. Cooper, B.~I. Halperin, C.-K. Hu, and I.~M. Ruzin, Phys. Rev. B {\bf 55},
   4551  (1997).

\bibitem{RuzCH96}
I.~M. Ruzin, N.~R. Cooper, and B.~I. Halperin, Phys. Rev. B {\bf 53},  1558
  (1996).

\bibitem{Shi99}
E. Shimshoni, Phys. Rev. B {\bf 60},  10691  (1999).

\bibitem{ShiAK98}
E. Shimshoni, A. Auerbach, and A. Kapitulnik, Phys. Rev. Lett. {\bf 80},  3352
  (1998).

\bibitem{SimH94}
S.~H. Simon and B.~I. Halperin, Phys. Rev. Lett. {\bf 73},  3278  (1994).

\bibitem{JanMZ99}
M. Janssen, M. Metzler, and M.~R. Zirnbauer, Phys. Rev. B {\bf 59},  15836
  (1999).

\bibitem{KagHA95}
V. Kagalovsky, B. Horovitz, and Y. Avishai, Phys. Rev. B {\bf 52},  R17044
  (1995).

\bibitem{KagHA97}
V. Kagalovsky, B. Horovitz, and Y. Avishai, Phys. Rev. B {\bf 55},  7761
  (1997).

\bibitem{KleM95}
R. Klesse and M. Metzler, Europhys. Lett. {\bf 32},  229  (1995).

\bibitem{KleZ01}
R. Klesse and M.~R. Zirnbauer, Phys. Rev. Lett. {\bf 86},  2094  (2001),
  {ArXiv}: cond-mat/0010005.

\bibitem{LeeC94}
D.~K.~K. Lee and J.~T. Chalker, Phys. Rev. Lett. {\bf 72},  1510  (1994).

\bibitem{LeeCK94}
D.~K.~K. Lee, J.~T. Chalker, and D.~Y.~K. Ko, Phys. Rev. B {\bf 50},  5272
  (1994).

\bibitem{RuzF95}
I. Ruzin and S. Feng, Phys. Rev. Lett. {\bf 74},  154  (1995).

\bibitem{WanLW94}
Z. Wang, D.-H. Lee, and X.-G. Wen, Phys. Rev. Lett. {\bf 72},  2454  (1994).

\bibitem{Ior82}
S.~V. Iordanskii, Solid State Commun. {\bf 43},  1  (1982).

\bibitem{StaA95}
D. Stauffer and A. Aharony, {\em Perkolationstheorie} (VCH, Weinheim, 1995).

\bibitem{KocHKP91a}
S. Koch, R.~J. Haug, K.~v. Klitzing, and K. Ploog, Phys. Rev. B {\bf 43},  6828
   (1991).

\bibitem{JanMMW98}
M. Janssen, R. Merkt, J. Meyer, and A. Weymer, Physica {\bf 256--258},  65
  (1998).

\bibitem{HucK90}
B. Huckestein and B. Kramer, Phys. Rev. Lett. {\bf 64},  1437  (1990).

\bibitem{HuoB92}
Y. Huo and R.~N. Bhatt, Phys. Rev. Lett. {\bf 68},  1375  (1992).

\bibitem{ButILP85}
M. {B\"{u}ttiker}, Y. Imry, R. Landauer, and S. Pinhas, Phys. Rev. B {\bf 31},
  6207  (1985).

\bibitem{Sha82}
B. Shapiro, Phys. Rev. Lett. {\bf 48},  823  (1982).

\bibitem{ChaRKH00}
J.~T. Chalker {\it et~al.}, Phys. Rev. B {\bf 65},  012506  (2002), {ArXiv}:
  cond-mat/0009463.

\bibitem{FreJM98}
P. Freche, M. Janssen, and R. Merkt,  in {\em Proceedings of the Ninth
  International Conference on Recent Progress in Many Body Theories}, edited by
  D. Neilson (World Scientific, Singapore, 1998), {ArXiv}: cond-mat/9710297.

\bibitem{FreJM99}
P. Freche, M. Janssen, and R. Merkt, Phys. Rev. Lett. {\bf 82},  149  (1999).

\bibitem{Jan98}
M. Janssen, Phys. Rep. {\bf 295},  1  (1998).

\bibitem{KagHAC99}
V. Kagalovsky, B. Horovitz, Y. Avishai, and J.~T. Chalker, Phys. Rev. Lett.
  {\bf 82},  3516  (1999).

\bibitem{MerJH98}
R. Merkt, M. Janssen, and B. Huckestein, Phys. Rev. B {\bf 58},  4394  (1998).

\bibitem{GruRS97}
I.~A. Gruzberg, N. Read, and S. Sachdev, Phys. Rev. B {\bf 55},  10593  (1997).

\bibitem{HoC96}
C.-M. Ho and J.~T. Chalker, Phys. Rev. B {\bf 54},  8708  (1996), {ArXiv}:
  cond-mat/9605073.

\bibitem{Kim96}
Y.~B. Kim, Phys. Rev. B {\bf 53},  16420  (1996).

\bibitem{KonM97}
J. Kondev and J.~B. Marston, Nucl. Phys. B {\bf 497},  639  (1997), {ArXiv}:
  cond-mat/9612223.

\bibitem{Lee94}
D.-H. Lee, Phys. Rev. B {\bf 50},  10788  (1994).

\bibitem{LudFSG94}
A.~W.~W. Ludwig, M.~P.~A. Fisher, R. Shankar, and G. Grinstein, Phys. Rev. B
  {\bf 50},  7526  (1994).

\bibitem{MarT99}
J.~B. Marston and S. Tsai, Phys. Rev. Lett. {\bf 82},  4906  (1999).

\bibitem{Zir94}
M.~R. Zirnbauer, {Ann. Phys. (Leipzig)} {\bf 3},  513  (1994).

\bibitem{Zir97}
M.~R. Zirnbauer, J. Math. Phys. {\bf 38},  2007  (1997), {ArXiv}:
  cond-mat/9701024.

\bibitem{Ber78}
J. Bernasconi, Phys. Rev. B {\bf 18},  2185  (1978).

\bibitem{ReyKS77}
P.~J. Reynolds, W. Klein, and H.~E. Stanley, J. Phys. C {\bf 10},  L167
  (1977).

\bibitem{StaA92}
D. Stauffer and A. Aharony, {\em Introduction to Percolation Theory} (Taylor
  and Francis, London, 1992).

\bibitem{CaiRSR01}
P. Cain, R.~A. {R\"{o}mer}, M. Schreiber, and M.~E. Raikh, Phys. Rev. B {\bf
  64},  235326  (2001), {ArXiv}: cond-mat/0104045.

\bibitem{JanMW98}
M. Janssen, R. Merkt, and A. Weymer, {Ann. Phys. (Leipzig)} {\bf 7},  353
  (1998).

\bibitem{WeyJ98}
A. Weymer and M. Janssen, {Ann. Phys. (Leipzig)} {\bf 7},  159  (1998),
  {ArXiv}: cond-mat/9805063.

\bibitem{CaiRR03}
P. Cain, R.~A. {R\"{o}mer}, and M.~E. Raikh, Phys. Rev. B {\bf 67},  075307
  (2003), {ArXiv}: cond-mat/0209356.

\bibitem{Meh91}
M.~L. Mehta, {\em Random Matrices and the Statistical Theory of Energy Levels}
  (Academic Press, New York, 1991).

\bibitem{PreFTV92}
W.~H. Press, B.~P. Flannery, S.~A. Teukolsky, and W.~T. Vetterling, {\em
  Numerical Recipes in {FORTRAN}}, 2nd ed. (Cambridge University Press,
  Cambridge, 1992).

\bibitem{Fer88}
H.~A. Fertig, Phys. Rev. B {\bf 38},  996  (1988).

\bibitem{BeeH91}
C.~W.~J. Beenakker and H. van Hoiuten,  in {\em Solid State Physics: Advances
  in Research and Applications}, edited by H. Ehrenreich and D. Turnbull
  (Academic, San Diego, 1991), Vol.~44, p.\ 207.

\bibitem{MacK81}
A. MacKinnon and B. Kramer, Phys. Rev. Lett. {\bf 47},  1546  (1981).

\bibitem{FisL81}
D.~S. Fisher and P.~A. Lee, Phys. Rev. B {\bf 23},  6851  (1981).

\bibitem{ChoF97}
S. Cho and M.~P.~A. Fisher, Phys. Rev. B {\bf 55},  1637  (1997).

\bibitem{JovW98}
B. Jovanovic and Z. Wang, Phys. Rev. Lett. {\bf 81},  2767  (1998).

\bibitem{WanJL96}
Z. Wang, B. Jovanovic, and D.-H. Lee, Phys. Rev. Lett. {\bf 77},  4426  (1996).

\bibitem{WanLS98}
X. Wang, Q. Li, and C.~M. Soukoulis, Phys. Rev. B {\bf 58},  3576  (1998).

\bibitem{AviBB99}
Y. Avishai, Y. Band, and D. Brown, Phys. Rev. B {\bf 60},  8992  (1999).

\bibitem{Wig51}
E.~P. Wigner, Proc. Camb. Phil. Soc. {\bf 47},  790  (1951).

\bibitem{Haa92}
F. Haake, {\em Quantum Signatures of Chaos}, 2nd ed. (Springer, Berlin, 1992).

\bibitem{PotS98}
H. Potempa and L. Schweitzer, J. Phys.: Condens. Matter {\bf 10},  L431
  (1998), {ArXiv}: cond-mat/9804312.

\bibitem{BraMP98}
D. Braun, G. Montambaux, and M. Pascaud, Phys. Rev. Lett. {\bf 81},  1062
  (1998), {ArXiv}: cond-mat/9712256.

\bibitem{SchP98}
L. Schweitzer and H. Potempa, Physica A {\bf 266},  486  (1998), {ArXiv}:
  cond-mat/9809248.

\bibitem{FyoM97}
Y.~V. Fyodorov and A.~D. Mirlin, Phys. Rev. B {\bf 55},  16001  (1997).

\bibitem{ShkSSL93}
B.~I. Shklovskii {\it et~al.}, Phys. Rev. B {\bf 47},  11487  (1993).

\bibitem{BatSK98}
M. Batsch, L. Schweitzer, and B. Kramer, Physica B {\bf 249},  792  (1998),
  {ArXiv}: cond-mat/9710011.

\bibitem{BatSZK96}
M. Batsch, L. Schweitzer, I.~K. Zharekeshev, and B. Kramer, Phys. Rev. Lett.
  {\bf 77},  1552  (1996), {ArXiv}: cond-mat/9607070.

\bibitem{FeiAB95}
M. Feingold, Y. Avishai, and R. Berkovits, Phys. Rev. B {\bf 52},  8400
  (1995), {ArXiv}: cond-mat/9503058.

\bibitem{KawOSO96}
T. Kawarabayashi, T. Ohtsuki, K. Slevin, and Y. Ono, Phys. Rev. Lett. {\bf 77},
   3593  (1996), {ArXiv}: cond-mat/9609226.

\bibitem{Met99}
M. Metzler, J. Phys. Soc. Japan {\bf 68},  144  (1999).

\bibitem{MetV98}
M. Metzler and I. Varga, J. Phys. Soc. Japan {\bf 67},  1856  (1998).

\bibitem{OhtO95}
T. Ohtsuki and Y. Ono, J. Phys. Soc. Japan {\bf 64},  4088  (1995), {ArXiv}:
  cond-mat/9509146.

\bibitem{OnoOK96}
Y. Ono, T. Ohtsuki, and B. Kramer, J. Phys. Soc. Japan {\bf 65},  1734  (1996),
  {ArXiv}: cond-mat/9603099.

\bibitem{ZhaK97}
I.~K. Zharekeshev and B. Kramer, Phys. Rev. Lett. {\bf 79},  717  (1997),
  {ArXiv}: cond-mat/9706255.

\bibitem{AbrALR79}
E. Abrahams, P.~W. Anderson, D.~C. Licciardello, and T.~V. Ramakrishnan, Phys.
  Rev. Lett. {\bf 42},  673  (1979).

\bibitem{HofS94b}
E. Hofstetter and M. Schreiber, Phys. Rev. B {\bf 49},  14726  (1994), {ArXiv}:
  cond-mat/9402093.

\bibitem{ZhaK95b}
I.~K. Zharekeshev and B. Kramer, Jpn. J. Appl. Phys. {\bf 34},  4361  (1995),
  {ArXiv}: cond-mat/9506114.

\bibitem{ZhaK95c}
I.~K. Zharekeshev and B. Kramer, Phys. Rev. B {\bf 51},  17239  (1995).

\bibitem{HofS93}
E. Hofstetter and M. Schreiber, Phys. Rev. B {\bf 48},  16979  (1993).

\bibitem{SleO99a}
K. Slevin and T. Ohtsuki, Phys. Rev. Lett. {\bf 82},  382  (1999), {ArXiv}:
  cond-mat/9812065.

\bibitem{Cai04}
P. Cain, Ph.D. thesis, Technische {Universit\"{a}t} Chemnitz, 2004.

\bibitem{ShaHLT98}
D. Shahar {\it et~al.}, Solid State Commun. {\bf 107},  19  (1998), {ArXiv}:
  cond-mat/9706045.

\bibitem{WeiH83}
A. Weinrib and B.~I. Halperin, Phys. Rev. B {\bf 27},  413  (1983).

\bibitem{Wei84}
A. Weinrib, Phys. Rev. B {\bf 29},  387  (1984).

\bibitem{SanMK03}
N. Sandler, H.~R. Maei, and J. Kondev, Phys. Rev. B {\bf 68},  205315  (2003),
  {ArXiv}: cond-mat/0304616.

\bibitem{WeiLTP92}
H.~P. Wei, S.~Y. Lin, D.~C. Tsui, and A.~M.~M. Pruisken, Phys. Rev. B {\bf 45},
   3926  (1992).

\bibitem{KocHKP92}
S. Koch, R.~J. Haug, K.~v. Klitzing, and K. Ploog, Phys. Rev. B {\bf 46},  1596
   (1992).

\bibitem{Lan00}
G. Landwehr, private communication.

\bibitem{KocHKP91b}
S. Koch, R.~J. Haug, K.~v. Klitzing, and K. Ploog, Phys. Rev. Lett. {\bf 67},
  883  (1991).

\bibitem{CobBF99}
D.~H. Cobden, C.~H.~W. Barnes, and C.~J.~B. Ford, Phys. Rev. Lett. {\bf 82},
  4695  (1999).

\bibitem{CobK96}
D.~H. Cobden and E. Kogan, Phys. Rev. B {\bf 54},  R17316  (1996).

\bibitem{EngSKT93}
L.~W. Engel, D. Shahar, C. Kurdak, and D.~C. Tsui, Phys. Rev. Lett. {\bf 71},
  2638  (1993).

\bibitem{She94}
D.~L. Shepelyansky, Phys. Rev. Lett. {\bf 73},  2607  (1994).

\bibitem{ApaR03}
V. Apalkov and M.~E. Raikh, Phys. Rev. B {\bf 68},  195312  (2003), arXiv:
  cond-mat/0303170.

\end{thebibliography}
\end{document}